\documentclass[revtex4]{emulateapjold}

\usepackage{color}
\usepackage{wasysym}

\usepackage{amsmath}

\newcommand{\mytilde}{\raise.17ex\hbox{$\scriptstyle\sim$}}
\newcommand{\brk}[1]{$[$#1$]$}
\newcommand{\angbrk}[1]{$\langle$#1$\rangle$}
\newcommand{\kepler}{{\it Kepler}}

\newcommand{\Rsun}{${R_\odot}$}

\newcommand{\Mear}{${M_\oplus}$}
\newcommand{\Rear}{${R_\oplus}$}

\newcommand{\Rstar}{${R_\star}$}
\newcommand{\Teff}{${T_{\rm eff}}$}

\newcommand{\Logg}{$\log{\rm g}$}

\newcommand{\Porb}{$P_{\rm orb}$}
\newcommand{\Mp}{$M_{\rm p}$}
\newcommand{\Rp}{$R_{\rm p}$}
\newcommand{\Kp}{$K_{\rm p}$}
\newcommand{\cjb}{}

\slugcomment{Submitted}

\begin{document}
\title{TERRESTRIAL PLANET OCCURRENCE RATES FOR THE {\it KEPLER} GK~DWARF SAMPLE}

\author{
Christopher~J.~Burke\altaffilmark{1},
Jessie~L.~Christiansen\altaffilmark{2},
F.~Mullally\altaffilmark{1},
Shawn Seader\altaffilmark{1},
Daniel Huber\altaffilmark{3,4,5},
Jason~F.~Rowe\altaffilmark{1},
Jeffrey~L.~Coughlin\altaffilmark{1},
Susan~E.~Thompson\altaffilmark{1},
Joseph Catanzarite\altaffilmark{1},
Bruce~D.~Clarke\altaffilmark{1},
Timothy~D.~Morton\altaffilmark{6},
Douglas~A.~Caldwell\altaffilmark{1},
Stephen~T.~Bryson\altaffilmark{7},
Michael~R.~Haas\altaffilmark{7},
Natalie~M.~Batalha\altaffilmark{7},
Jon~M.~Jenkins\altaffilmark{7},
Peter Tenenbaum\altaffilmark{1},
Joseph~D.~Twicken\altaffilmark{1},
Jie Li\altaffilmark{1},
Elisa Quintana\altaffilmark{1},
Thomas Barclay\altaffilmark{8},
Christopher~E.~Henze\altaffilmark{7},
William~J.~Borucki\altaffilmark{7},
Steve~B.~Howell\altaffilmark{7},
Martin Still\altaffilmark{8}
}

\email{christopher.j.burke@nasa.gov}
\altaffiltext{1}{SETI Institute/NASA Ames Research Center, Moffett Field, CA 94035, USA}
\altaffiltext{2}{NASA Exoplanet Science Instititute, California Institute of Technology, Pasadena, CA 91125, USA}
\altaffiltext{3}{Sydney Institute for Astronomy (SIfA), School of Physics, University of Sydney, NSW 2006, Australia}
\altaffiltext{4}{SETI Institute, 189 Bernardo Avenue, Mountain View, CA 94043, USA}
\altaffiltext{5}{Stellar Astrophysics Centre, Department of Physics and Astronomy, Aarhus University, Ny Munkegade 120, DK-8000 Aarhus C, Denmark}
\altaffiltext{6}{Department of Astrophysics, Princeton University, Princeton, NJ, 08544, USA}
\altaffiltext{7}{NASA Ames Research Center, Moffett Field, CA 94035, USA}
\altaffiltext{8}{BAERI/NASA Ames Research Center, Moffett Field, CA 94035, USA}
\begin{abstract}

We measure planet occurrence rates using the planet candidates
discovered by the Q1-Q16 \kepler\ pipeline search.  This study
examines planet occurrence rates for the \kepler\ GK dwarf target
sample for planet radii, 0.75$\leq$\Rp$\leq$2.5~\Rear, and orbital
periods, 50$\leq$\Porb$\leq$300~days, with an emphasis on a thorough
exploration and identification of the most important sources of
systematic uncertainties.  Integrating over this parameter space, we
measure an occurrence rate of $F_{0}$=0.77 planets per star, with an
allowed range of 0.3$\leq F_{0}\leq$1.9.  {\cjb The allowed range takes
  into account both statistical and systematic uncertainties, and
  values of $F_{0}$ beyond the allowed range are significantly in
  disagreement with our analysis}.  We generally find higher planet
occurrence rates and a steeper increase in planet occurrence rates
towards small planets than previous studies of the \kepler\ GK~dwarf
sample.  Through extrapolation, we find that the one year orbital
period terrestrial planet occurrence rate, $\zeta_{1.0}$=0.1, with an
allowed range of 0.01$\leq\zeta_{1.0}\leq$2, where $\zeta_{1.0}$ is
defined as the number of planets per star within 20\% of the \Rp\ and
\Porb\ of Earth.  For G~dwarf hosts, the $\zeta_{1.0}$ parameter space
is a subset of the larger $\eta_{\oplus}$ parameter space, thus
$\zeta_{1.0}$ places a lower limit on $\eta_{\oplus}$ for G~dwarf
hosts.  From our analysis, we identify the leading sources of
systematics impacting \kepler\ occurrence rate determinations as:
reliability of the planet candidate sample, planet radii, pipeline
completeness, and stellar parameters.

\end{abstract}


\section{Introduction}\label{sec:intro}

The \kepler\ data set is the only currently available experiment
capable of detecting and characterizing the planetary content of the
Milky Way down to the regime of terrestrial planets orbiting within 1~AU
of solar-type stars \citep{BOR10,KOC10}.  The ongoing detailed analysis of
\kepler\ data can provide accurate and precise measurement of the
occurrence rate of Earth analogs beyond the Solar System, which is a
critical parameter for understanding the potential for life outside
the Solar System \citep{DRA13,PRA13} and a quantity of immense
interest across the disciplines of science, engineering, philosophy, and
sociology.

\kepler\ builds upon the enormous efforts of the astronomical
community in
filling out the parameter space of planet properties in numerous
stellar environments
\citep{FIS05,MAR05,NAE05,GOU06,SAH06,CUM08,BOW10,HOW10,BAY11,MAY11,VAN11,WRI12,BON13,MEI13,CLA14}. \kepler\ expands
the planetary discovery space to terrestrial planets within 1~AU of
solar-type stars.  \kepler\ data enables the study and simulations of
planetary formation to finally be confronted with their predictive
outcomes in the regime of rocky planets \citep{IDA04,BEN14,MOR15}.
In addition, \kepler\ constraints on the prevalence of rocky planets
in the habitable zone (HZ) of nearby stars provides a key input
in defining the scope of future missions that will probe the atmospheres of
extrasolar planets \citep{DRE13,KOU14,BAT14,LEG15,STA15}.

A substantial shortcoming for planet occurrence rate determinations using
the \kepler\ pipeline planet candidate samples
\citep{BOR11A,BOR11B,BAT13,BUR14,ROW15,MULL15} has been the unavailability of an
accurate model for the completeness of the \kepler\ pipeline \citep{BAT14}.
Previous planet occurrence determinations from \kepler\ data have
dealt with this shortcoming by employing simplifying assumptions as to
the pipeline completeness: assume the theoretical performance of the
Transiting Planet Search algorithm \citep[TPS][]{JEN02} with a
signal-to-noise ratio (SNR) threshold of 7.1 \citep{BOR11B}, designate
a higher SNR level where the planet sample is close to 100\% complete
\citep{CAT11,YOU11,HOW12,TRA12,DRE13,DON13,SIL15}, or simultaneously solve for a
parameterized completeness model in addition to planet occurrence
\citep{FRE13,MUL15,FAR14}.  Others avoid this shortcoming altogether
through an independent planet search pipeline and pipeline
completeness measurement \citep{PET13A,PET13B,DRE15}.

\citet{CHR15} rectify this shortcoming by directly measuring the
\kepler\ pipeline completeness of the Q1-Q16 \kepler\ pipeline run
\citep{TEN14} through Monte-Carlo transit injection and recovery
tests.  In this study, we make use of the \citet{CHR15}
\kepler\ pipeline completeness parameterization in order to derive the
planet occurrence rates from the resulting Q1-Q16 \kepler\ planet
candidate sample of \citet{MULL15}.  {\cjb Another highlight of this
  study is a comprehensive analysis of the systematic errors present
  in deriving planet occurrence rates with \kepler\ data.  As
  exemplified in \citet{YOU11} and \citet{DON13}, we undertake a
  sensitivity analysis where we iteratively change an input assumption
  and recalculate the occurrence rates.  We investigate the following
  input assumptions: pipeline completeness systematics, orbital
  eccentricity, stellar parameter systematics, planet parameter
  systematics, and planet sample classification systematics.}

This paper is organized as follows.  Section~\ref{sec:compmod}
describes the pipeline completeness model that quantifies the survey
completeness for any target observed by \kepler.
Sections~\ref{sec:stars} and~\ref{sec:planets} summarize the stellar
properties and planet sample from the Q1-Q16 \kepler\ pipeline run
adopted for derivation of the planet occurrence rates.  We extend the
analysis techniques of \citet{YOU11} by increasing the complexity of
the parameterized model for the planet occurrence rate and employ
Markov Chain Monte-Carlo (MCMC) methods for solving the parameter
estimation problem in Section~\ref{sec:method}.
Section~\ref{sec:baseline} presents results for the planet occurrence
rate using a baseline set of inputs, and we
thoroughly explore the systematic errors in this result through a
sensitivity analysis in Section~\ref{sec:sensit}.  We compare the
occurrence rate analysis with previous efforts in
Section~\ref{sec:disc}.  We apply the resulting occurrence rates to
determine the occurrence rate for terrestrial planets with an orbital period equivalent to Venus in Section~\ref{sec:venus} as well
as extrapolating these results toward longer periods
(Section~\ref{sec:extrap}) in order to measure a one year terrestrial planet
occurrence rate in Section~\ref{sec:earth}.  Finally, Section~\ref{sec:conclusion} summarizes the future work necessary to improve
the accuracy for the resulting planet occurrence rates.

\section{\kepler\ Pipeline Completeness Model}\label{sec:compmod}

This section details an analytic star-by-star model for the
\kepler\ pipeline completeness.  A critical component for modeling the
completeness of \kepler\ observations is simulating the performance of
the TPS pipeline module which is responsible for characterizing the
noise present in a light curve and detection of the transit signals
\citep{JEN02,TEN12,TEN13,TEN14}.  The performance of a transit survey
can be fully specified with intensive, end-to-end Monte Carlo signal
injection and recovery tests \citep{WEL05,BUR06,HAR09,CHR13,SEA14}.
Unfortunately, due to their numerically intensive nature, Monte-Carlo
injection tests are not amenable to a systematic sensitivity analysis,
and the tests are limited to the subset of targets that one performs
the analysis upon.  Therefore, we present a simplified analytic model
for the \kepler\ pipeline that can be readily applied to any observed
\kepler\ target using a minimum of input data.  Fortunately, the joint
noise characterization, filtering, and detection properties of TPS
were designed to facilitate a well defined and tested detector
response for transit signals even in the presence of astrophysical
broad-band or red noise \citep{JEN02}.  Given the well defined
properties of the TPS detector, our analytic completeness model can
achieve high fidelity after it is calibrated with Monte-Carlo
injection tests.  For a single target, we parameterize the pipeline
completeness over a two-dimensional grid of orbital period, \Porb, and
planet radius, \Rp.

\subsection{Multiple Event Statistic Estimation}\label{sec:mes}

Modeling pipeline completeness requires modeling the statistical
behavior of TPS and its response to noise in the presence of a signal
\citep{JEN02,SEA13}.  In the presence of broad-band red noise, TPS
considers the so-called Multiple Event Statistic (MES) to measure the
strength of a potential transit signal.  In the null hypothesis case
of no signal present, the MES distribution is Gaussian with an average
of zero and unit variance.  In the alternative hypothesis case for the
presence of a signal, the MES distribution is Gaussian but the average
MES is shifted proportional to the SNR of the transit signal.  The
first step for modeling pipeline completeness is to estimate the
expected MES of a transit signal for a specified \Porb\ and \Rp.  This
requires an estimate of the expected transit duration,
\begin{equation}
\tau_{\rm dur}=4 \left( \frac{P_{\rm orb}}{\rm 1 day}\right)\left( \frac{R_\star}{a}\right)\sqrt{1-e^2}\; {\rm hr},\label{eq:trandur}
\end{equation}
where $e$ is the orbital eccentricity and the stellar radius, \Rstar,
and orbital semi-major axis, $a$, are in a consistent set of units.
In Equation~(\ref{eq:trandur}), we assume \Rp$\ll$\Rstar, shorten the
transit duration from the central crossing time by a factor of $\pi/4$
for its expectation assuming a uniform distribution of $\cos{i}$ for
the orbital inclination \citep{GIL00,SEA03}, and include the expected
dependence on the transit duration with $e$ \citep{BUR08}.  We explore
the sensitivity of our results to $e>0$ in
Section~\ref{sec:sensitecc}.

Next, we determine the noise present in the light curve data averaged
over the transit duration of interest.  TPS estimates the time varying
noise present in a light curve, the so-called Combined Differential
Photometric Precision \citep[CDPP,][]{KOC10,CHR12}.  CDPP varies with time
and is calculated over the same 14 transit durations, $\tau_{\rm dur,
  srch}=[1.5, 2.0, 2.5, 3.0, 3.5, 4.5, 5.0,\allowbreak 6.0, 7.5, 9.0, 10.5, 12.0, 12.5, 15.0]\,
{\rm hr}$, that are used in the transiting planet search \citep[see
  Figure~3 of][]{CHR12}.  For the analytic completeness model, we
employ a summary statistic, a robust root-mean-square of the CDPP
(robCDPP), for the light curve noise.  In testing it was found that
the non-robust root-mean-square CDPP (rmsCDPP) statistic typically
reported by the Q1-Q16 pipeline data products (SOC build 9.2 and
earlier) can be biased.  The
bias in the rmsCDPP arises when the distribution formed from the CDPP
time series values is asymmetric with the outlying tail of the
distribution inordinately affecting the results.

We calculate robCDPP by replacing the typical components of the root-mean-square that includes the mean/dc component
\begin{equation}
x_{\rm rms}^2=\bar{x}^2+\sigma^2,
\end{equation}
where $\bar{x}$ is the mean and $\sigma$ is the standard deviation, with their robust equivalents
\begin{equation}
x_{\rm rob,rms}^2={\rm median}(x)^2+(1.4826\times{\rm MAD})^2,
\end{equation}
where MAD is the median absolute deviation.  The robCDPP is calculated
after removing both invalid and deweighted data identified in the same
manner as during the transit search.  In order to remove the influence
of strong signals in estimating the noise present in a light curve,
the robCDPP value adopted for the completeness model is calculated on
the light curve after all the potential transit signatures have been
removed in the Data Validation (DV) multiple planet search
\citep{WU10}.  On average, the robCDPP to rmsCDPP ratio is 1.03
with a sample standard deviation of 0.06.  A higher fidelity model of
pipeline completeness would employ the full CDPP time series.
However, in tests we find that when the number of expected transits
contributing to a detection, $N_{\rm trn}\gtrsim 5$, the robCDPP
summary is sufficient to model pipeline completeness rather than a
more time-consuming calculation involving the full CDPP time series
(see Section~\ref{sec:compexample}).

For a given $\tau_{\rm dur}$, we interpolate within the grid of 14
 robCDPP values to estimate the noise for that
duration, $\sigma_{\rm cdpp}$.  For values of $\tau_{\rm dur}$ outside
the $\tau_{\rm dur, srch}$ grid, we adopt the end point robCDPP for
$\sigma_{\rm cdpp}$.  Alternative extrapolation methods such as
assuming a $\sigma_{\rm cdpp}\propto 1/\sqrt{\tau_{\rm dur}}$ dependence
or linear extrapolation were not stable for the sometimes complicated
behavior of robCDPP as a function of $\tau_{\rm dur}$.

The next step is to estimate the expected transit signal depth,
$\Delta$.  Since the TPS search algorithm employs a square box-car
signal template match to the signal for detection, the appropriate
$\Delta$ is the average signal depth over $\tau_{\rm dur}$ rather than
the purely geometric depth $\delta=k^2=(R_{\rm p}/R_\star)^2$, where
$k$ is the physical radius ratio.  From the results of the transit
injection study of \citet{CHR15}, we determine that for limb darkened
transit signals, on average $\Delta=0.84\Delta_{\rm max}$, where
$\Delta_{\rm max}$ is the depth at closest approach or maximum transit
depth.  When averaged over a uniform distribution of impact parameter,
$b$, we find using the limb darkened transit model of \citet{GIM06}
that $\Delta_{\rm max}/k^2=c+s\, k$ is well fit by a linear
relationship with parameters $c$ and $s$ that vary with the limb
darkening profile of the stellar intensity.  When using a linear limb
darkening law, $I=1-u(1-\cos{\theta})$, where $I$ is the stellar
intensity relative to the stellar center, $u$ is the linear
coefficient, and $\theta$ is the line-of-sight stellar-surface-normal
angle, with a coefficient appropriate for G dwarfs, $u=0.6$, we
determine best fit values of ($c=1.0874$, $s=1.0187$).  We note that
$c$ and $s$ are weakly dependent upon limb darkening, taking on values
($c=1.0696$, $s=1.001$) for $u=0.5$ and ($c=1.1068$, $s=1.0379$)
for $u=0.7$.  The needed value of $\Delta$ is expressible in terms of
$k$ using the previous equations to provide the single transit event
SNR, $\Delta/\sigma_{\rm cdpp}$.

The MES is calculated by averaging the transit signal strength over
multiple transit events.  The resulting MES=$\sqrt{N_{\rm
    trn}}\Delta/\sigma_{\rm cdpp}$, where $N_{\rm trn}=(T_{\rm
  obs}/P_{\rm orb})\times f_{\rm duty}$ is the expected number of
transit events, $T_{\rm obs}$ is the time baseline of observational
coverage for a target and $f_{\rm duty}$ is the observing duty cycle.
The observing duty cycle, $f_{\rm duty}$, is defined as the fraction
of $T_{\rm obs}$ with valid observations.  The
\kepler\ spacecraft experiences planned data gaps for data download
and other spacecraft operations.  These data gaps result in an overall
duty cycle of $\sim 95$\% for targets observed for all quarters.  In
addition, the duty cycle is suppressed further in the transit search
to $\sim 88$\% by a set of data weights applied in TPS.  Data near
gaps suffers from spacecraft systematics, thus TPS deweights data near
gaps using a smooth exponential decay functional form that
goes from fully deweighted data at the gap edge to full data weighting
over a span of 2~days.  We calculate $f_{\rm duty}$ by dividing the
number of cadences with an overall deweighting factor $>0.5$ by the
total number of cadences within $T_{\rm obs}$.  In addition, we force a
floor of $N_{\rm trn}\geq 3$ in the MES estimate since TPS requires at
least 3 transit events for detection.  $T_{\rm obs}$ and $f_{\rm
  duty}$ are calculated in the initial call to TPS before transit
signals have been identified and removed by DV.

\subsection{Pipeline Completeness Modeling}\label{sec:sensfunc}

The TPS search algorithm design results in a well-defined pipeline
completeness that is, in the limit of broadband noise, a function of
MES alone \citep{JEN02}.  Since the TPS detection statistic, MES, is
distributed as a Gaussian with unit variance, the pipeline
completeness (fraction of transit signals present in the data that are
recovered by the pipeline) has a theoretical expected form of
\begin{equation}
P_{\rm det}{\cjb (\rm MES)}=\frac{1}{2}+\frac{1}{2}{\rm erf}\left[ \frac{({\rm MES}-{\rm MES}_{\rm thresh})}{\sqrt{2}} \right],\label{eq:tpsthy}
\end{equation}
where ${\rm MES_{\rm thresh}}=7.1$ is the adopted detection threshold
\citep{JEN02B,JEN02}.  However, the presence of stochastic, impulsive
systematics of instrumental or astrophysical origin in the time series
that are not due to a transit signal results in an overwhelming number
of false alarm detections when MES is the only criteria
employed for detection.  In the Q1-Q16 pipeline run, 57\% of targets
result in a detection when based upon MES alone \citep{TEN14}.  To
mitigate the false positive detections, additional criteria (or vetoes)
are employed that quantify how consistent the depths, shape, and
duration of individual events are with each other
\citep{TEN13,TEN14,SEA13}.  The additional vetoes cause the pipeline
completeness to be suppressed relative to the theoretical expectation
given in Equation~(\ref{eq:tpsthy}).

\citet{CHR15} quantify the resulting suppression of the pipeline
completeness through Monte-Carlo transit injection and recovery tests.
They find that the gamma cumulative distribution function (CDF)
provides a good fit to the suppressed pipeline completeness,
\begin{equation}
P_{\rm gamma}(x|a,b)=\frac{1}{b^a\Gamma(a)}\int^{x}_{0}t^{a-1}\exp^{-t/b}dt,
\end{equation}
where $\Gamma(a)$ is the gamma function {\cjb and the argument to the gamma
CDF, $x={\rm MES}-4.1-({\rm MES}_{\rm thresh}-7.1)$, is related to MES
by an offset of 4.1 in order to achieve a good fit}.  The parameters
for the gamma CDF adopted for this study are $a$=4.65 and $b$=0.98.
In rare cases, due to the timeout limits of the TPS planet search,
MES$_{\rm thresh}$ is higher than the normal MES$_{\rm thresh}$=7.1.
Section~\ref{sec:mes} can be used to provide a mapping for the 2D grid
of \Rp\ and \Porb\ onto MES.  The pipeline detection efficiency
provides a mapping from the MES to the pipeline completeness.

\subsection{Window Function}\label{sec:windowfunc}

The final component of the pipeline completeness model accounts for
the limits of the data coverage for meeting the planet search
detection requirement of having at least $N_{\rm tr}\geq 3$.  The
transit survey window function, $P_{\rm win}$, quantifies the
probability that a requisite number of transits required for detection
occurs in the observational data \citep{GAU00,VON09,BUR14B}.  Since
\kepler\ operates in the high duty cycle regime, we adopt the binomial
analytic window function as discussed in \citet{BUR14B} and originally
introduced by \citet{DEE04}.  The analytic window function matches the
average behavior of the fully numerical window function \citep[see
  Figure~9 of][]{BUR14B}, and requires as input $T_{\rm obs}$, \Porb,
and $f_{\rm duty}$.  {\cjb Following Appendix~A.4 of \citet{BUR14B}, the window function probability of detecting at least three transits can be explicitly written out in the binomial approximation as
\begin{multline}
P_{{\rm win}, \geq 3}=1-(1-f_{\rm duty})^{M}-M\, f_{\rm duty}(1-f_{\rm duty})^{M-1} \\
   -\frac{M(M-1)}{2}\, f_{\rm duty}^{2}(1-f_{\rm duty})^{M-2}, 
\end{multline}
where $M=T_{\rm obs}/$\Porb.  }
The final completeness model results from $P_{\rm comp}=P_{\rm win}\times P_{\rm det}$.

{\cjb For targets with data in all Q1-Q16 quarters, the analytic
  window function predicts $P_{\rm win, \geq 3}\geq 0.98$ for
\Porb$\leq$300~days.  We compare the impact on $P_{\rm comp}$ between
using the analytic window function and a full numerical window
function in Section~\ref{sec:compexample}.}

\subsection{Worked Example and Limitations}\label{sec:compexample}

\begin{figure}
\includegraphics[trim=0.25in 0.05in 0.35in 0.3in,scale=0.5,clip=true]{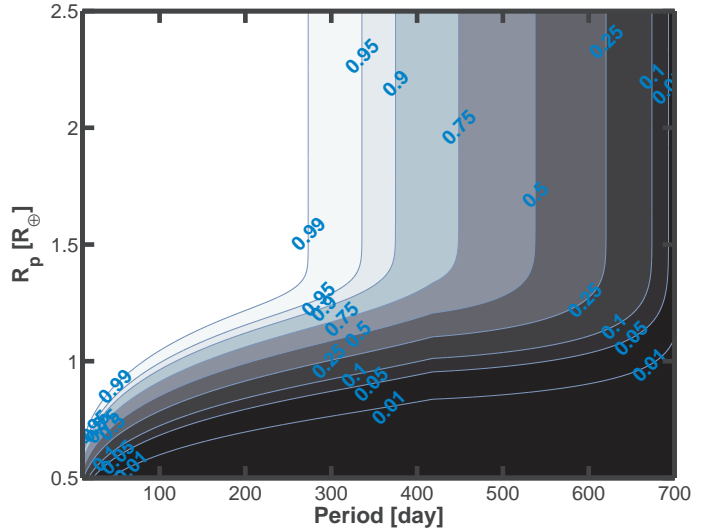}
\caption{
Fractional completeness model for the host to Kepler-22b (KIC: 10593626) in the Q1-Q16 pipeline run using the analytic model described in Section~\ref{sec:compmod}.\label{fig:exampfig}}
\end{figure}

As a worked example, we demonstrate the calculation of $P_{\rm comp}$
for the host to Kepler-22b \citep{BOR12}.  This target with Kepler
Input Catalog (KIC) identifier 10593626 has stellar parameters
\Rstar=0.98~\Rsun, \Teff=5640~K, \Logg=4.44 as compiled by the Q1-Q16
stellar catalog of \citet{HUB14}.  In order to generate $P_{\rm comp}$
over the two-dimensional space of \Rp\ and \Porb\ shown in
Figure~\ref{fig:exampfig}, we employ the following input values:
$f_{\rm duty}=0.879$, $T_{\rm obs}=1426.7$~days, $e=0$, ${\rm MES_{\rm
    thresh}}=7.1$, and the 14 robCDPP noise estimates, $\sigma_{\rm
  cdpp,
  14}=[36.2, 33.2, 31.0, 29.4, 28.0, 26.1, 25.4,\allowbreak 24.2, 23.1, 22.4, 21.9, 21.8, 21.7, 21.5]$
ppm.  The quantities necessary to generate $P_{\rm comp}$ for all
\kepler\ targets searched for planets in the Q1-Q16 pipeline run are
available as part of the Q1-Q16 \kepler\ stellar table hosted by the
NASA Exoplanet
Archive.\footnote{http://exoplanetarchive.ipac.caltech.edu}

\begin{figure}
\includegraphics[trim=0.25in 0.05in 0.35in 0.3in,scale=0.5,clip=true]{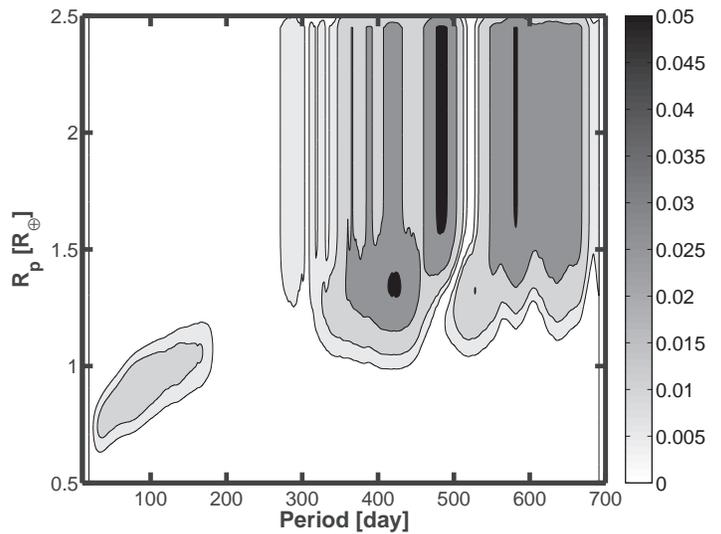}
\caption{
Absolute difference for pipeline completeness between the analytic model described in Section~\ref{sec:compmod} and a more accurate numerical pipeline completeness model that employs the full CDPP time series and numerical window function for the KIC target 10593626.\label{fig:diffmodel}}
\end{figure}

In order to investigate potential biases in the analytic pipeline
completeness model, we show the absolute difference between the
analytic model presented in this study and a higher fidelity
completeness model that is available for future pipeline runs in
Figure~\ref{fig:diffmodel}.  {\cjb The higher fidelity completeness
  model replaces two components of the completeness model described in
  Section~\ref{sec:compmod}.  First, the analytic window function
  approximation is replaced by a full numerical window function
  calculated during TPS to take into account data gaps and data
  deweighting consistent with the planet search.  Second, the
  simplified MES estimate of Section~\ref{sec:mes} that employs the
  robCDPP values is replaced by the ``1-$\sigma$ depth function''
  (1SDF).  For the 1SDF, TPS quantifies the transit signal depth that
  yields a MES=1 as a function of \Porb\ for all 14 transit durations
  searched taking into account the full details of the full CDPP time
  series and deweighted data.  }

Differences between the completeness estimates are largest ($\sim
0.06$) towards longer periods (\Porb$\geq$ 300~days).  {\cjb The occurrence
rate is $\propto P_{\rm comp}^{-1}$ \citep{YOU11}, thus errors in the
pipeline completeness can propagate directly to occurrence rates.
Future study is needed in order to characterize the net impact of the
higher fidelity completeness model on occurrence rates for a full
sample of \kepler\ targets.}

Although this initial test affirms the efficacy of the analytic
completeness model for this well behaved \kepler\ target, this
comparison does not fully test all its simplifying assumptions.  The
above test does not check the accuracy of our assumption of a simple
dependence of the pipeline completeness on MES alone and the adopted
$\Delta$ to $k$ conversion.  Due to a single injection per target,
\citet{CHR15} cannot rule out the possibility that the pipeline
completeness may depend on additional parameters beyond MES and
contain a strong star-by-star dependence.  Results for small samples or
individual targets may systematically differ from the average pipeline
completeness results of \citet{CHR15}.  However, \citet{CHR15} have
characterized the average pipeline completeness as a function of MES
when averaged over a large sample of targets as is the case in this study.
Future studies will focus on characterizing the star-by-star
dependence of pipeline completeness by comparing the simplified
completeness model outlined in this study to the higher fidelity
completeness model using the full numerical window function and
1$\sigma$ depth function for larger samples of targets.  In addition,
we are implementing support for $\sim 10^4$ transit injections on a
single target employing the NASA Ames Pleiades super computing
facility \citep{SEA14}.

An additional shortcoming of any pipeline completeness model is
the inescapable dependence on the assumed stellar parameters,
eccentricity, and stellar binarity.  Stellar parameters and
eccentricity are employed in defining $\tau_{\rm dur}$ which sets the
time scale relevant for the integrated noise level, and stellar
binarity can result in third light contamination that impacts the
assumed planet radius. In this study, we treat the pipeline
completeness as having no uncertainty due to the incomplete
understanding of the stellar parameters, eccentricity, and stellar
binarity.  However, we do explore sensitivity in the derived planet
occurrence rates to alternative assumptions for the stellar parameters
and non-zero eccentricities in Section~\ref{sec:sensit}.

{\cjb Although it is a distinct process separate from the pipeline
  generation of TCEs, the vetting classification process of TCEs into
  planet candidates and false positives also shapes the overall
  completeness of the planet candidate sample, and the vetting relied
  upon on a manual classification (i.e., human inspection) procedure
  \citep{MULL15}.  The TCE vetting process has an unquantified false
  negative rate of incorrectly classifying valid planet candidates as
  false positives and unconscious human biases and/or errors.  For
  this study, we assume the vetting process is 100\% complete,
  unbiased, and correct.  However, we do investigate the sensitivity
  of our results on this assumption in Section~\ref{sec:sensitplanets}
  by varying the planet candidate sample.  }

The analytic pipeline completeness model and input data presented in
this study are only relevant to the TCE population generated by the
Q1-Q16 pipeline run \citep{TEN14} using the SOC 9.1 software release.
The next TCE release \citep{SEA15} used the SOC 9.2 software, which
introduced changes to the data analysis and planet search algorithm
that influences the pipeline sensitivity.  In SOC 9.2, TPS implemented
a bootstrap noise characterization algorithm during the search in
order to recalibrate the detection threshold \citep{SEA15}.  The
bootstrap noise characterization test allows the effective MES
threshold to be a function of \Porb\ as opposed to being independent
of \Porb\ in the SOC 9.1 software release.  In addition, the
box-car signal template matched to the data in TPS was replaced by an
average limb-darkened signal template to yield a better signal match.
However, changing the match template influences the noise statistics
(such as robCDPP) and the $\Delta$ to $k$ conversion.  Finally, the
relation between pipeline completeness and MES is being analyzed
through Monte-Carlo transit injection using the updated software
release.

\section{Stellar Properties}\label{sec:stars}

Precise and accurate planet occurrence rates depend on precise and
accurate stellar properties.  Stellar properties influence three areas
relevant to planet occurrence rates: the measured planet radii, the
estimated transit duration, and geometric transit probability.  In
this study, we adopt stellar parameters from the Q1-Q16 KIC revision
of \citet{HUB14}.  \citet{HUB14} update \kepler\ target stellar
parameters by adopting literature values and additional observations
(asteroseismology, spectroscopy, and photometry) that have become
available since the original KIC observations \citep{BRO11}.  With an
improved observational database, \citet{HUB14} derive stellar
parameters by fitting the observations to isochrones from the
Dartmouth Stellar Evolution Database \citep{DOT08} using a $\chi^2$
minimization.

\begin{figure}
\includegraphics[trim=0.25in 0.05in 0.35in 0.3in,scale=0.5,clip=true]{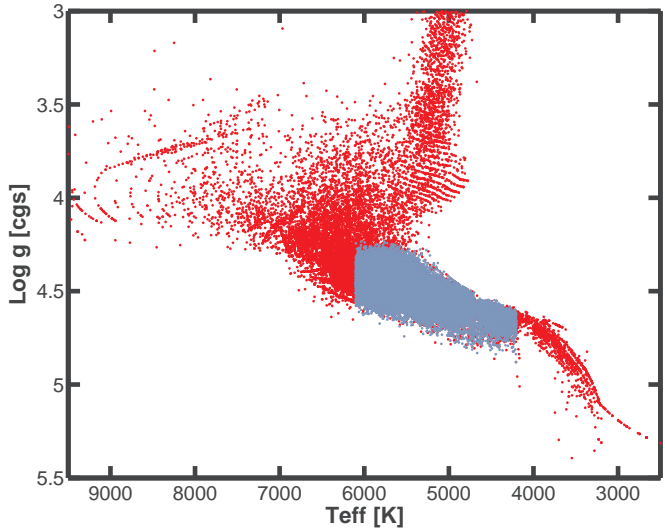}
\caption{Stellar \Teff\ as a function of \Logg\ for the observed \kepler\ targets from the Q1-Q16 stellar catalog of \citet{HUB14} (red points), and the GK dwarf targets selected for this study (gray points).  For efficient plotting only a randomly selected subsample of the full catalog is shown.\label{fig:loggteff}}
\end{figure}

For this study we focus on planet occurrence for the G and K dwarf
sample observed by \kepler.  Previous occurrence rate calculations
indicate significant variations in the planet population as a function
of stellar \Teff\ for the \kepler\ sample \citep{HOW12,MUL15,BUR15}.
In order to simplify the planet occurrence model by avoiding the
dependence on stellar \Teff, we focus on the GK dwarfs rather than the
full FGKM dwarf sample.  \citet{BUR15} find that the planet occurrence
rates agree when the G and K dwarf \kepler\ targets are analyzed
separately in an \Rp\ and \Porb\ parameter space similar to this
study.  We select G and K dwarfs by making the following cuts on
the stellar parameter catalog of \citet{HUB14}:
4200$\leq$\Teff$\leq$6100 K and \Rstar$\leq$ 1.15 \Rsun.  In addition,
we focus on the \kepler\ targets with nearly continuous coverage over
the entire Q1-Q16 data span of the mission.  We select targets with an
observation data spanning $T_{\rm obs}>$ 2 yr, duty cycle $f_{\rm
  duty}>$ 0.6, and a robCDPP$\leq$ 1000 ppm at the 7.5 hr transit
duration.  The lower limit on $f_{\rm duty}$ ensures the inclusion of
the targets that are impacted by the CCD electronics loss in Q4 of the
\kepler\ mission \citep{BAT13}.  The above cuts result in 91,567
targets in our sample.  Figure~\ref{fig:loggteff} shows \Teff\ and
\Logg\ for the full catalog of \citet{HUB14} (red points) along with
the \kepler\ targets selected using the above criteria (gray points).
Table~\ref{tab:kicused} provides the \kepler\ identification number
and a binary flag to indicate that the target belongs to the baseline
stellar sample selected by the above criteria.  Adopting \Teff=5200~K
as the dividing line separating the G and K dwarfs, 80\% of our
stellar sample belong to the G dwarf category.

\begin{figure}
\includegraphics[trim=0.25in 0.05in 0.35in 0.3in,scale=0.5,clip=true]{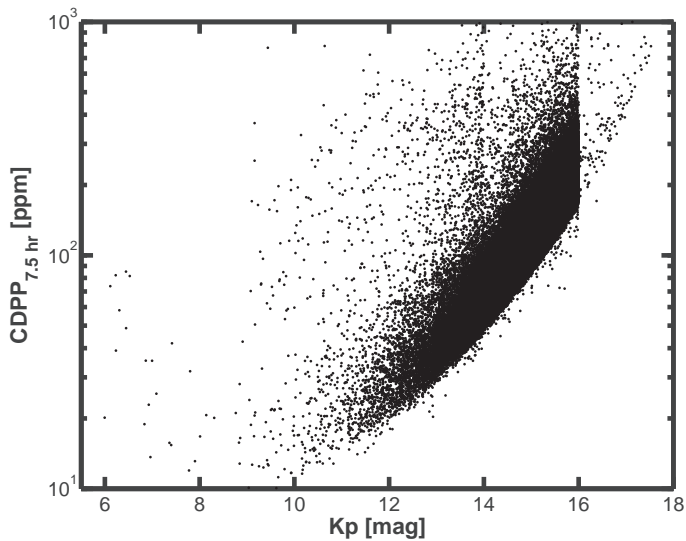}
\caption{Distribution of the robCDPP at the 7.5 hr time scale as a function of \Kp\ for the G and K dwarfs analyzed in this study.  Mean properties of the target stars as a function of \Kp\ are provided in Table~\ref{tab:magbinprops}.\label{fig:noisekpmag}}
\end{figure}

\begin{figure}
\includegraphics[trim=0.25in 0.05in 0.35in 0.3in,scale=0.5,clip=true]{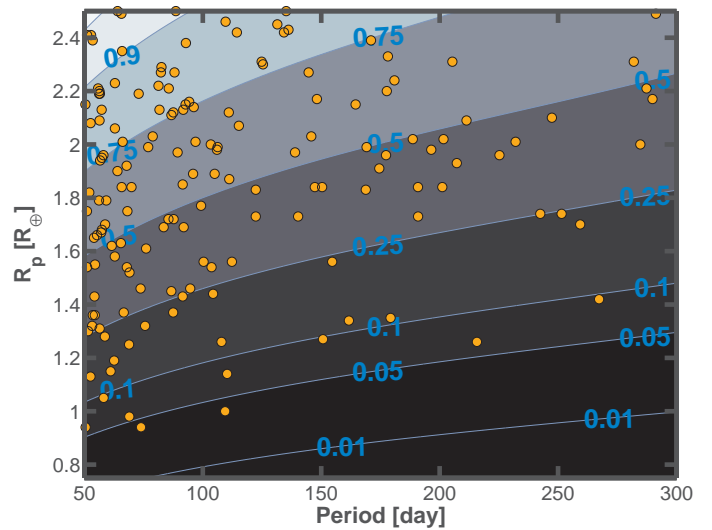}
\caption{Pipeline completeness model, $P_{\rm comp}$, averaged over the GK dwarf sample is shown by the contour levels over the \Porb\ and \Rp\ parameter space.  The Q1-Q16 \kepler\ planet candidate sample of \citet{MULL15} found around the GK dwarf sample is also shown (orange points).\label{fig:avgdet}}
\end{figure}

In Figure~\ref{fig:noisekpmag} we present the noise distribution of
the selected GK dwarfs as a function of \Kp.  \citet{GIL11} and
\citet{CHR12} discuss the instrumental and astrophysical sources of
noise in the \kepler\ data.  We also provide mean properties of the GK
dwarfs as a function of \Kp\ in Table~\ref{tab:magbinprops}.  In
magnitude wide bins, Table~\ref{tab:magbinprops} provides the bin
centers, number of targets, mean stellar properties (\Rstar, \Logg,
and \Teff), and the 7.5 hr robCDPP for the 10$^{\rm th}$, 50$^{\rm
  th}$, and 90$^{\rm th}$ percentiles in each bin.  Using the pipeline
completeness model of Section~\ref{sec:compmod}, we show in
Figure~\ref{fig:avgdet} the average pipeline completeness for the GK
dwarf sample in terms of detection probability contour levels over the
\Porb\ and \Rp\ parameter space examined in this study.  The figure
does not include the effects of the geometric probability to transit.
We discuss our choice of \Porb\ and \Rp\ space examined for this study
in Section~\ref{sec:planets}.

\begin{deluxetable}{@{\hspace{4 pt}}c|@{\hspace{4 pt}}c|@{\hspace{4 pt}}c}
\tablewidth{0pt}
\tablecaption{{\rm \kepler\ GK Dwarf Target Samples}}
\startdata
\hline
\hline
KIC ID & Baseline Sample & Original KIC Sample \\
\hline
   757450 & 1 & 1 \\
   891901 & 0 & 1 \\
   891916 & 1 & 1 \\
   892718 & 1 & 1 \\
   892772 & 1 & 1 \\
   892832 & 1 & 1 \\
   892834 & 1 & 1 \\
   892882 & 1 & 1 \\
   892911 & 1 & 1 \\
   892946 & 0 & 1 \\
\hline
\enddata
\label{tab:kicused}
\tablecomments{This table is available in its entirety in a machine-readable form in the online journal.  A portion is shown here for guidance regarding its form and content.}
\end{deluxetable}

\begin{deluxetable}{@{\hspace{4 pt}}c@{\hspace{4 pt}}c@{\hspace{4 pt}}c@{\hspace{4 pt}}c@{\hspace{4 pt}}c@{\hspace{4 pt}}c@{\hspace{4 pt}}c@{\hspace{4 pt}}c}
\tablewidth{0pt}
\tablecaption{{\rm Kepler Target Summary}}
\startdata
\hline
\hline
\Kp  & N & \angbrk{\Rstar} & \angbrk{\Logg} & \angbrk{\Teff} & robCDPP$_{7.5 \rm hr}$ & robCDPP$_{7.5\rm hr}$ & robCDPP$_{7.5\rm hr}$ \\
\brk{mag} & & \brk{\Rsun} & \brk{cgs} & \brk{K} & 10$^{\rm th}$\% \brk{ppm} & 50$^{\rm th}$\% \brk{ppm} & 90$^{\rm th}$\% \brk{ppm} \\
\hline
 8 &    9 &  0.93 &  4.48 & 5658 &     9.3 &    13.1 &    27.4 \\
 9 &   36 &  0.90 &  4.48 & 5543 &    12.5 &    22.6 &   169.9 \\
10 &  117 &  0.93 &  4.46 & 5640 &    13.1 &    23.9 &   169.8 \\
11 &  364 &  0.91 &  4.47 & 5608 &    17.5 &    26.8 &   142.5 \\
12 & 1312 &  0.89 &  4.48 & 5613 &    22.6 &    32.1 &    67.6 \\
13 & 4964 &  0.90 &  4.48 & 5643 &    32.8 &    44.9 &    78.2 \\
14 & 16011 &  0.88 &  4.50 & 5625 &    51.3 &    70.9 &   106.8 \\
15 & 41649 &  0.86 &  4.52 & 5563 &    88.2 &   123.9 &   180.8 \\
16 & 27027 &  0.81 &  4.56 & 5413 &   147.8 &   193.4 &   271.4 \\
17 &   58 &  0.82 &  4.54 & 5201 &   328.0 &   456.0 &   788.9 \\
\hline
\enddata
\label{tab:magbinprops}
\end{deluxetable}

\section{Planet Properties}\label{sec:planets}

We measure the planet occurrence rate using the Q1-Q16 pipeline run
\citep{TEN14} and the resulting \kepler\ planet candidate sample from
\citet{MULL15}.  We choose to limit our analysis to
50$\leq$\Porb$\leq$300~days and rocky to mini-Neptune planets with
0.75$\leq$\Rp$\leq$2.5~\Rear.  The \Porb\ range under investigation
has several advantages.  \citet{MULL15} classified the new and
pre-existing KOIs into planet candidate and false positives for the
\Porb$>$50~days regime.  Thus, the selected period range represents a
uniform classification of planet candidates following the procedures
of \citet{MULL15}.  The cumulative KOI catalog for \Porb$<$50~days
currently consists of classifications from \citet{BAT13},
\citet{BUR14}, \citet{ROW15}, and \citet{MULL15}.  Toward shorter and
longer \Porb, the population of instrumental false alarm detections
increases rapidly \citep[see bottom panel of Figure 4 in ][]{TEN14}.
The vetting process employed by \citet{MULL15} effectively removes much of
the instrumental false alarm contamination, but the remaining
contamination is potentially higher outside this \Porb\ range.

For a majority of \kepler\ targets, \Porb$\sim$300~days is roughly the
transition between planet candidates having at least 4-5 transit
events contributing to the detection and planet candidates having the
minimum three transit events contributing to the detection \citep[see
  Figure~9 of][]{BUR14B}.  The three transit event
(\Porb$\gtrsim$300~days) low MES planet candidates are the most
challenging candidates to vet properly \citep{MULL15} and further work
is needed to understand the false alarm rate of this population.
Thus, we exclude them from planet occurrence rate calculations for the
time being.  Third, approximating the behavior of the full CDPP time
series by the summary robCDPP statistics for the pipeline completeness
model can be inaccurate at long periods when there is an increased
chance that the few transit events available can occur during outliers
of the CDPP noise distribution.

The astrophysical false positive contamination rates have
observationally \citep{SAN12,DES15} and theoretically
\citep{MOR11,FRE13} been shown to increase towards shorter \Porb.
Also for shorter periods, \Porb$<$3~days, the harmonic removal filter
in the pipeline increasingly removes transiting planet signals
\citep{CHR13,CHR15}.  The detection efficiency reported in
\citet{CHR15} is calculated for \Porb$>$10 days.  Thus, the detection
efficiency does not take into account the impact of the harmonic
removal filter.
 
{\cjb In order to select the \kepler\ planet candidate sample for
  analysis, we must limit the KOI planet candidates to ones recovered
  in the Q1-Q16 pipeline run.  The analysis of \citet{MULL15}
  uniformly vetted the pre-existing KOIs and new KOIs that made a
  Threshold Crossing Event (TCE) corresponding to the KOI ephemeris
  for \Porb$\geq$50 days.  To supplement the list of planet
  candidates, we make special exceptions for systems with strong
  transit timing variations (TTV).  The pipeline only searches for
  transits with a uniform ephemeris and targets with high SNR and
  strong TTVs can result in multiple TCEs at the incorrect period, but
  corresponding to one to a few of the high SNR transit events.  If it
  is clear that the TCE corresponds to one or a few of the single
  events of a TTV system, but formally the TCE ephemeris does not
  match the KOI ephemeris, then that TTV KOI planet candidate is
  accepted as recovered by the pipeline and included in our analysis.
  We provide the KOI planet candidates deemed as recovered by the
  Q1-Q16 pipeline run in Table~\ref{tab:koiused}.  The recovered KOIs
  within the 50$\leq$\Porb$\leq$300~days limits for the baseline
  occurrence rate calculation are indicated by a binary flag in
  Table~\ref{tab:koiused}.}

{\cjb In addition, we provide KOI planet
  candidates recovered in the Q1-Q16 pipeline run in the
  10$\leq$\Porb$\leq$50~day range in order to support analysis of the
  \kepler\ planet candidate sample outside the parameter space of this
  study.  The 10$\leq$\Porb$\leq$50~day range has not been uniformly
  vetted, so for KOI recovery designations in this parameter space we
  start with the cumulative KOI table combining results from all the
  \kepler\ planet catalogs
  \citep{BOR11A,BOR11B,BAT13,BUR14,ROW15,MULL15}.  We employ an
  ephemeris matching routine \citep{MULL15} to judge whether a
  Threshold Crossing Event (TCE) detection from the pipeline run
  \citep{TEN14} matches the ephemeris for the KOIs.  KOIs with a match
  statistic satisfying that the KOI ephemeris overlaps with $\geq
  90$\% of the TCE's transit events are automatically accepted as
  recovered.  The KOI ephemeris from a previous catalog may not be
  accurate enough to guarantee 100\% of the transit overlapping, and
  the 90\% matching level is sufficient to automatically adopt the TCE
  as corresponding to the KOI without manual inspection.  Also,
  special exception was made for systems with strong TTVs as discussed
  above.  Table~\ref{tab:koiused} lists the KOIs from the cumulative
  KOI list that are designated as being recovered in the Q1-Q16
  pipeline run for \Porb$>$10~days.}

\begin{deluxetable}{l|@{\hspace{4 pt}}c|@{\hspace{4 pt}}c|@{\hspace{4 pt}}c|@{\hspace{4 pt}}c|@{\hspace{4 pt}}c}
\tablewidth{0pt}
\tablecaption{{\rm Q1-Q16 Planet Candidate Samples}}
\startdata
\hline
\hline
KOI number & Baseline & Original KIC & High Reliability & Full Long Period & Trimmed Long Period \\
\hline
  12.01 & 0 & 0 & 0 & 0 & 0 \\
  41.01 & 0 & 0 & 0 & 0 & 0 \\
  41.03 & 0 & 0 & 0 & 0 & 0 \\
  42.01 & 0 & 0 & 0 & 0 & 0 \\
  51.01 & 0 & 0 & 0 & 0 & 0 \\
  70.01 & 0 & 0 & 0 & 0 & 0 \\
  70.03 & 0 & 1 & 0 & 0 & 0 \\
  70.05 & 0 & 0 & 0 & 0 & 0 \\
  72.02 & 0 & 0 & 0 & 0 & 0 \\
  75.01 & 0 & 0 & 0 & 0 & 0 \\
\hline
\enddata
\label{tab:koiused}
\tablecomments{This table is available in its entirety in a machine-readable form in the online journal.  A portion is shown here for guidance regarding its form and content.}
\end{deluxetable}

For the planet radii, we adopt estimates from the uniform KOI analysis
in \citet{ROW15} and \citet{MULL15}.  The technique is described fully
in \citet{ROW14}.  Briefly, the flux time series data are detrended
with a moving cubic polynomial fit.  Data occurring during transit or
near gaps are excluded from the moving polynomial fit.  Assuming a
circular orbit, fixed limb darkening parameters from \citet{CLA11},
and stellar parameters from \citet{HUB14}, \citet{ROW14} use a
Markov-chain Monte-Carlo methodology to estimate the best
fitting parameters of a limb darkened transit model.

Overall, we find 156 planet candidates orbit stars in the GK dwarf
sample within the \Porb, \Rp\ parameter space under investigation.  We
illustrate the planet candidate sample (orange points) in
Figure~\ref{fig:avgdet}.  In our analysis described in
Section~\ref{sec:method}, we do not take into account the
uncertainties on \Rp.  However, we do explore the influence that
systematic changes to the planet candidate sample, stellar sample, and
independent model fits have on the resulting planet occurrence rates
in Section~\ref{sec:sensit}.

In this study, we do not model or include the impact of astrophysical
false positive contamination in our sample.  Following the process
outlined in \citet{MOR12}, a preliminary astrophysical false positive
analysis was completed for 108 (70\%) of the baseline planet candidate
sample.  We find that the average and median false positive
probabilities for the calculated sample are 4\% and 0.6\%,
respectively.  Twelve planet candidates in the sample have an
astrophysical false positive probability $p_{\rm fpp}\geq\ 10\%$ and
the highest is 60\%.  The astrophysical false positive
  contamination for the parameter space under investigation is within
  the statistical and systematic uncertainties and can be safely
  ignored for this study.  However, for shorter and longer \Porb, the
  astrophysical false positive contamination becomes increasingly
  important.

\section{Planet Occurrence Rate Method}\label{sec:method}

In order to infer the underlying planet occurrence rate from the
observed distribution of \kepler\ planet candidates, we extend the
methodology of \citet{YOU11}.  \citet{YOU11} present a parametric
model for the planet distribution function (PLDF) and use likelihood
maximization techniques to estimate the parameters that best describe
the observed planet candidate distribution and the parameter
uncertainties.  We extend the method of \citet{YOU11} by employing
Bayesian parameter estimation theory using Markov Chain Monte-Carlo
(MCMC) methods to numerically evaluate the posterior distribution of
the PLDF parameters \citep{GRE05}.  We were motivated to replace the
intuitive and analytic minimization method of \citet{YOU11} with a
Bayesian MCMC parameter estimation method in order to analyze a more
complicated PLDF model and enable future efforts to explore higher
dimensional models including, for example, dependence on stellar
parameters.

Following \citet{YOU11}, we employ the Poisson distribution for the
likelihood.  A helpful description motivating the Poisson likelihood
is given in Section~5.3.2 of \citet{LOR92}, {\cjb and the Poisson
  likelihood is commonly used in order to interpret astronomical
  detections with a varying survey sensitivity
  \citep{TAB02,ALL07,CUM08,KRA12,FOR14,NIE14}}.  Also, the point
process statistics literature \citep{DAL88,BAD07} rigorously shows
that the Poisson likelihood is appropriate for analyses of spatial
point data.  For this application, the observed planet candidate distribution is
treated as an inhomogeneous Poisson process where the PLDF describes
the dependence of the Poisson process intensity on \Porb\ and \Rp.  

{\cjb Independent of the choice of likelihood, one is free to choose any parametric form for the PLDF model.  Previous work has indicated that a power law form of the PLDF describes the \kepler\ observations \citep{YOU11,HOW12,DON13} over portions of the \Rp\ and \Porb\ parameter space}.  For this study, we adopt a PLDF dependent upon \Porb\ and \Rp\ parameterized as a power law in \Porb\ and a broken power law in \Rp\ over a specified domain $P_{\rm min}\le P_{\rm orb} \le P_{\rm max}$ and $R_{\rm min} \le R_{\rm p} \le R_{\rm max}$:
\begin{multline}
\frac{{\rm d}^{2}f}{{\rm d}P_{\rm orb}{\rm d}R_{\rm p}}=F_{0}\, C_{\rm n}\, g(P_{\rm orb},R_{\rm p}) \\ 
=\left\{
\begin{array}{lr}
F_{0}\, C_{\rm n}\left( \frac{P_{\rm orb}}{P_{\rm o}}\right)^{\beta} \left( \frac{R_{\rm p}}{R_{\rm o}}\right)^{\alpha_{1}} & {\rm if}\, R_{\rm p} < R_{\rm brk} \\
F_{0}\, C_{\rm n}\left( \frac{P_{\rm orb}}{P_{\rm o}}\right)^{\beta} \left( \frac{R_{\rm p}}{R_{\rm o}}\right)^{\alpha_{2}} \left( \frac{R_{\rm brk}}{R_{\rm o}}\right)^{\alpha_{1}-\alpha_{2}} & {\rm if}\, R_{\rm p} \ge R_{\rm brk} 
\end{array}
\right. ,\label{eq:pldf}  
\end{multline}
where $F_{0}$ is the integrated planet occurrence rate, $C_{\rm n}$ is a normalization factor, $g({\bf x})$ is the shape function, $P_{\rm o}=(P_{\rm min}+P_{\rm max})/2$ and $R_{\rm o}=(R_{\rm min}+R_{\rm max})/2$ are domain scaling factors, $R_{\rm brk}$ is the break radius transition between the two \Rp\ power law exponents ($\alpha_{1}$ and $\alpha_{2}$), and $\beta$ is the \Porb\ power law exponent.  The $C_{\rm n}$ is determined from the normalization requirement,
\begin{equation}
\int\limits_{P_{\rm min}}^{P_{\rm max}}\int\limits_{R_{\rm min}}^{R_{\rm max}}\, C_{\rm n}\, g(P_{\rm orb},R_{\rm p}){\rm d}P_{\rm orb}{\rm d}R_{\rm p}=1.
\end{equation}
Overall, the PLDF has five free parameters: $F_{0}$, $\beta$, $\alpha_{1}$, $\alpha_{2}$, and $R_{\rm brk}$.  Following Equation~(18) of \citet{YOU11}, the Poisson likelihood of the data for a survey that detects $N_{\rm pl}$ planets around $N_{\star}$ survey targets is
\begin{equation}
L\propto\left[F_{0}^{N_{\rm pl}}\, C_{\rm n}^{N_{\rm pl}} \prod\limits_{i=1}^{N_{\rm pl}}\, g(P_{\rm orb},R_{\rm p})\right]\exp(-N_{\rm exp}), 
\end{equation}
where the PLDF model predicted number of detections from the survey is 
\begin{equation}
N_{\rm exp}=F_{0}\, C_{\rm n}\, \int\limits_{P_{\rm min}}^{P_{\rm max}}\int\limits_{R_{\rm min}}^{R_{\rm max}}\, \left[  \sum\limits_{j=1}^{N_{\star}}\,\eta_{j}(P_{\rm orb},R_{\rm p})\right]\, g(P_{\rm orb},R_{\rm p}){\rm d}P_{\rm orb}{\rm d}R_{\rm p},\label{eq:nexp}
\end{equation}
{\cjb and the likelihood ignores constant multiplicative factors.}
In Equation~(\ref{eq:nexp}), the underlying PLDF model is modified by the per-star transit survey effectiveness, $\eta_{j}({\bf x})$, summed over $N_{\star}$ targets in the sample, where $\eta_{j}({\bf x})=P_{j, {\rm comp}}\times P_{j, {\rm tr}}$ is the per-star pipeline completeness model of Section~\ref{sec:compmod} and $P_{j, {\rm tr}}$ is the geometric probability to transit.  The transit probability factor, $P_{j, {\rm tr}}=(R_{\star}/a)/(1-e^2)$, depends on the stellar parameters and orbital eccentricity \citep{BUR08}.

{\cjb The separable form between \Rp\ and \Porb\ of the PLDF adopted in this study, is overly restrictive if applied to a larger range of \Rp.  Previous studies have identified a dependence of the \Porb\ exponent, $\beta$, on planet radius \citep{DON13,FOR14}, with an apparent transition in the \Porb\ dependence around \Rp$\sim$4\Rear.  For the 0.75$\leq$\Rp$\leq$2.5~\Rear\ analysis region of this study we do not find evidence for a more complicated dependence between \Rp\ and \Porb\ being necessary based upon residuals between the observed and model fitted planet counts.  Also of note in Equation~(\ref{eq:nexp}), is that the summation of $\eta_{j}$ over the stellar sample is independent of the PLDF parameters.  Thus, the summation can be computed once for the analysis and the planet occurrence depends upon the integrated/average transit survey effectiveness alone rather than explicitly depending upon the per-star survey effectiveness.}

We complete the Bayesian posterior calculation by specifying uniform
priors for all parameters except for $F_{0}$ which has a prior that is
uniform in the logarithm.  The adopted MCMC implementation for this
analysis is based upon the description in \citet{GRE05} that employs a
Metropolis-Hastings algorithm with an automated iterative proposal
step-size determination and has been applied to transit model light
curve analysis \citep{BUR07}, transit timing analysis \citep{BUR10},
and radial velocity analysis \citep{CHA09,BAL11}.  For this study, we
do not incorporate uncertainty in \Rp and ignore contributions to the
planet candidate sample due to astrophysical and instrumental false
positives \citep[see][for a more in-depth discussion]{YOU11,MULL15}.
{\cjb In the case of multiple planet systems, adopting the Poisson
  likelihood treats multiple planets in a system as independent, and
  thus this method can not capture any structure and correlations
  between planet's in a single system, but captures the average
  behavior over a large sample of stars}.  However, we do constrain
the sensitivity of our results to these potential complications in
Section~\ref{sec:sensit}.

\section{Results}\label{sec:results}

In this section, we provide planet occurrence rate determinations based
upon the Q1-Q16 \kepler\ planet candidate sample of \citet{MULL15}
(see Section~\ref{sec:planets} and Table~\ref{tab:koiused}).  We focus
on the GK dwarf targets observed by \kepler\ using stellar parameters
from the catalog of \citet{HUB14} (see Section~\ref{sec:stars} and
Table~\ref{tab:kicused}).  We describe our analytic pipeline
completeness model in Section~\ref{sec:compmod} that employs the
pipeline detection efficiency as calibrated by the Monte-Carlo transit
injection and recovery provided by \citet{CHR15}.  The planet
occurrence rate is derived through a parameterized model for the PLDF,
where the parameters and their uncertainties are determined within a
Bayesian parameter estimation problem with the posterior estimated
through MCMC techniques (see Section~\ref{sec:method}).  The above set
of inputs represents our current best/baseline model for planet
occurrence rates, and we describe the results in
Section~\ref{sec:baseline}.  We then perform a sensitivity analysis in
Section~\ref{sec:sensit} in order to explore the systematic
uncertainty in the planet occurrence rates due to imprecise knowledge
of the baseline inputs.

\subsection{Baseline Results}\label{sec:baseline}

For the baseline result, we fit the PLDF over the parameter space of
0.75$<$\Rp$<$2.5 \Rear\ and 50$<$\Porb$<300$~days.  We tabulate 10,000
subsamples from the full MCMC posterior samples for all the parameters
along with the resulting likelihood and prior values in
Table~\ref{tab:baseposterior}.  The overall occurrence rate for this
parameter space $F_{0}=0.77\pm0.12$ planets per star.  Relying on the
statistical uncertainty alone, the 3-$\sigma$ upper limit $F_{0,
  3\sigma\, \rm U.L.}=1.3$ implies {\cjb that we cannot currently
  rule-out a scenario that when averaged over large samples of GK
  dwarfs there exists more planets in the analyzed parameter space
  than stellar hosts.  The 3-$\sigma$ lower limit $F_{0, 3\sigma\, \rm
    L.L}=0.49$ implies that for large samples of GK dwarfs there
  exists on average at least one planet in the analyzed parameter
  space for every two stellar hosts.}

\begin{deluxetable}{@{\hspace{4 pt}}c@{\hspace{4 pt}}c@{\hspace{4 pt}}c@{\hspace{4 pt}}c@{\hspace{4 pt}}c@{\hspace{4 pt}}c@{\hspace{4 pt}}c}
\tablewidth{0pt}
\tablecaption{{\rm PLDF Model Parameter Posterior Samples}}
\startdata
\hline
\hline
 $\alpha_{1}$ & $\alpha_{2}$ & $R_{\rm brk}$ & $\beta$ & $F_{0}$ & Ln(Likelihood) & Ln(Prior) \\
\hline
 -1.80587 &   9.60189 & 2.42398 &  -0.53218 &  1.04356 & -1154.1220 &  -11.3374 \\
 -0.91336 &  -7.31895 & 2.20004 &  -0.68832 &  0.74751 & -1152.1463 &  -11.3374 \\
 -2.55130 &  -1.50156 & 1.71089 &  -0.78946 &  1.04314 & -1156.0940 &  -11.3374 \\
 16.62721 &  -1.43560 & 0.91343 &  -0.68223 &  0.69378 & -1151.8822 &  -11.3374 \\
  6.77569 &  -1.25637 & 0.87505 &  -0.61621 &  0.67331 & -1153.4823 &  -11.3374 \\
 -1.34402 &  -7.41157 & 2.37924 &  -0.45503 &  0.93427 & -1155.0970 &  -11.3374 \\
 -1.90010 &  -3.45914 & 2.01354 &  -0.91872 &  0.97944 & -1155.2124 &  -11.3374 \\
  6.59736 &  -1.64782 & 1.00841 &  -0.55609 &  0.71677 & -1152.5281 &  -11.3374 \\
 -1.96796 &  -3.36621 & 2.29527 &  -0.34320 &  1.02223 & -1155.7010 &  -11.3374 \\
 16.05061 &  -2.14134 & 0.94454 &  -0.76077 &  0.83308 & -1152.5983 &  -11.3374 \\
\hline
\enddata
\label{tab:baseposterior}
\tablecomments{This table is available in its entirety in a machine-readable form in the online journal.  A portion is shown here for guidance regarding its form and content.}
\end{deluxetable}

\begin{figure}
\includegraphics[trim=0.05in 0.05in 0.35in 0.3in,scale=0.5,clip=true]{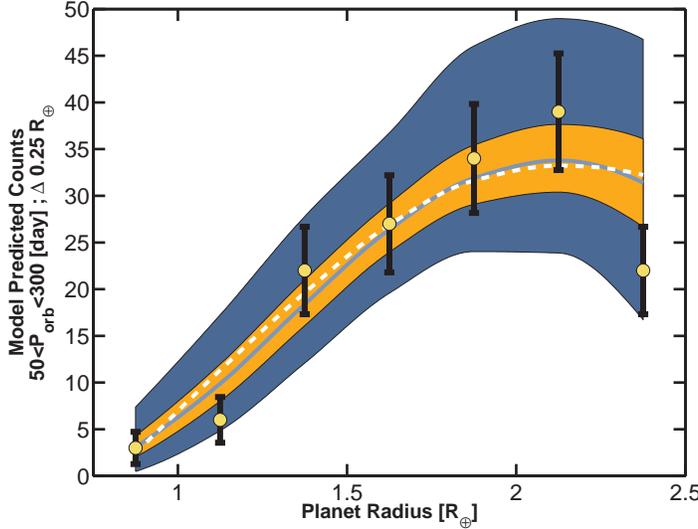}
\caption{{\cjb Comparison between the predicted planet sample from the planet occurrence rate model and the observed \kepler\ planet candidate sample.}  The observed, marginalized over 50$<$\Porb$<$300~days, histogram of \kepler\ planet candidate counts as a function of \Rp\ (points) can be compared to the maximum likelihood model for the predicted counts (white dash line).  Also shown is the posterior distributions of the model predicted counts for the median (blue solid line), 1-$\sigma$ region (orange region), and 3-$\sigma$ region (blue region) marginalized over \Porb\ and in bins of d\Rp=0.25 \Rear.
\label{fig:smallrp_rp_modelcomp}}
\end{figure}

\begin{figure}
\includegraphics[trim=0.05in 0.05in 0.35in 0.3in,scale=0.5,clip=true]{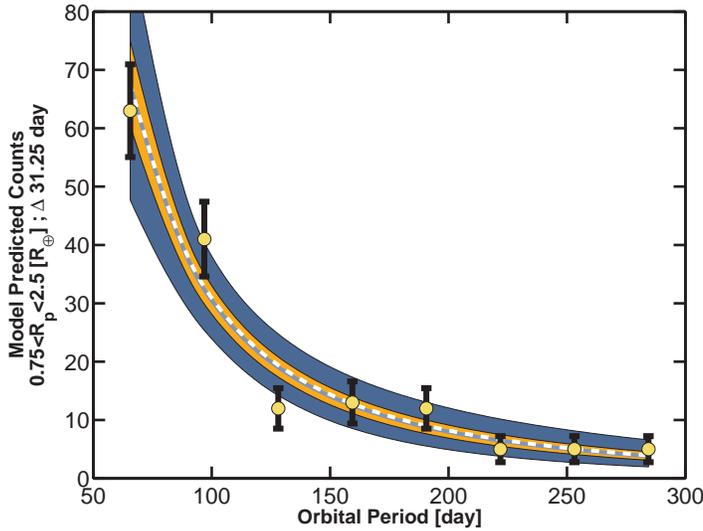}
\caption{Same as Figure~\ref{fig:smallrp_rp_modelcomp}, but marginalized over 0.75$<$\Rp$<$2.5 \Rear and bins of d\Porb=31.25~days.\label{fig:smallrp_per_modelcomp}}
\end{figure}

Figure~\ref{fig:smallrp_rp_modelcomp} shows how well the parametric
PLDF model {\cjb predicts} the observed, uncorrected \kepler\ planet
candidate counts summed over 50$<$\Porb$<$300~days in d\Rp=0.25
\Rear\ sized bins (points with uncertainties).  Evaluating the PLDF at
the parameters that maximize the likelihood fit to the data, we show
the model predicted counts ($N_{\rm exp}$ of Equation~(\ref{eq:nexp})
where the limits of integration are 50$<$\Porb$<$300~days and
d\Rp=0.25 \Rear) as the white dashed line.  In addition, we show the
median (solid blue line), 1-$\sigma$ (orange region), and 3-$\sigma$
(blue region) model predicted counts by evaluating $N_{\rm exp}$ using
10,000 random samples from the posterior PLDF parameter estimates from
the MCMC chain.  Figure~\ref{fig:smallrp_per_modelcomp} shows the
equivalent information, but along the \Porb\ dimension after
marginalizing over 0.75$<$\Rp$<$2.5 \Rear\ and d\Porb=31.25~days.
{\cjb The bin sizes for the abcissae in
  Figures~\ref{fig:smallrp_rp_modelcomp}~and~\ref{fig:smallrp_per_modelcomp}
  are chosen in order to balance segmenting the parameter space range
  into a high number of evenly sized bins and maintaining at least
  three detections in each bin.}

\begin{figure}
\includegraphics[trim=0.05in 0.05in 0.35in 0.3in,scale=0.5,clip=true]{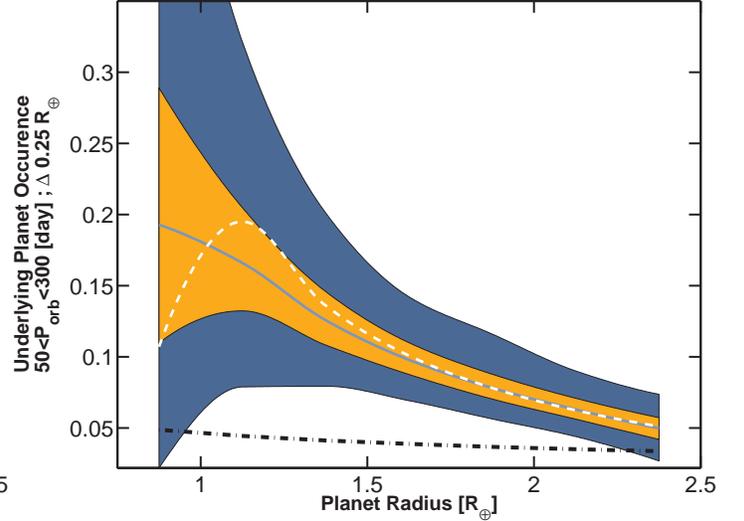}
\caption{Shows the underlying planet occurrence rate model.  Marginalized over 50$<$\Porb$<$300~days and bins of d\Rp=0.25~\Rear\ planet occurrence rates for the model parameters that maximize the likelihood (white dash line).  Posterior distribution for the underlying planet occurrence rate for the median (blue solid line), 1-$\sigma$ region (orange region), and 3-$\sigma$ region (blue region).  An approximate PLDF based upon results from \citet{PET13B} for comparison (dash dot line).\label{fig:smallrp_rp_underlying}}
\end{figure}

\begin{figure}
\includegraphics[trim=0.05in 0.05in 0.35in 0.3in,scale=0.5,clip=true]{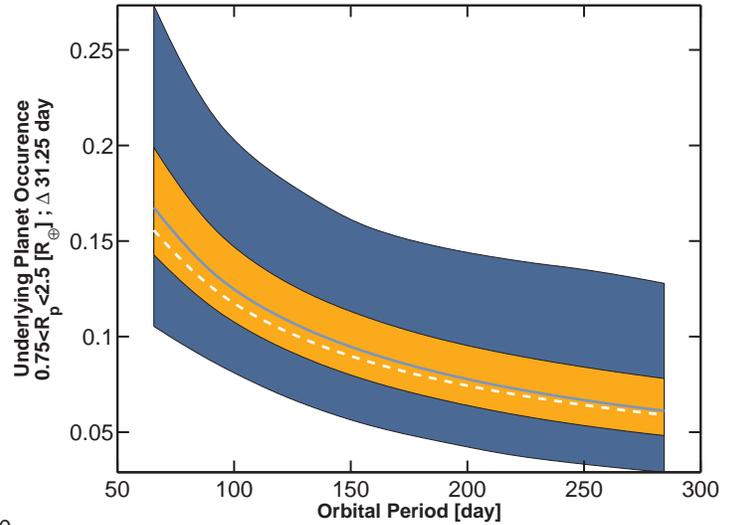}
\caption{Same as Figure~\ref{fig:smallrp_rp_underlying}, but marginalized over 0.75$<$\Rp$<$2.5 \Rear\ and bins of d\Porb=31.25~days.\label{fig:smallrp_per_underlying}}
\end{figure}

Figure~\ref{fig:smallrp_rp_underlying} quantifies the underlying PLDF
free of the deleterious effects of the \kepler\ pipeline completeness
and geometric transit probability.  The white dashed line,
representing the PLDF for parameters that maximize the likelihood of
the data, rises toward small planets with $\alpha_{2}=-1.8$ and has a
break near the edge of the parameter space.  Given the low numbers of
observed planet candidates in the smallest planet bins, the full
posterior allowed behavior (1-$\sigma$ orange region ; 3-$\sigma$ blue
region) cannot distinguish between a rising or falling PLDF for
\Rp$\lesssim1.5$ \Rear.  Figure~\ref{fig:smallrp_per_underlying} shows
the equivalent information, but along the \Porb\ dimension after
marginalizing over 0.75$<$\Rp$<$2.5 \Rear\ and d\Porb=31.25~days.

Formally, in our baseline analysis of the GK dwarf sample, the double
power law in the \Rp\ model is unwarranted relative to a single power
law according to the Bayesian information criterion (BIC) methodology
for model comparison.  However, we choose to provide the final results
in terms of the double power law model for the following reasons: (a)
The additional flexibility of the double power law model provides a
better fit to the smallest \Rp\ parameter space of most interest,
whereas the single power law model systematically overestimates (by
$\sim$0.5 $\sigma$ in a comparable data/model comparison to that shown
in Figure~\ref{fig:smallrp_rp_modelcomp}) the occurrence rates in the
smallest \Rp\ bins.  (b) The more complicated model ensures the
ability to adapt to variations in the PLDF in the
sensitivity analysis of Section~\ref{sec:sensit}.  (c) Previous work
on \kepler\ planet occurrence rates indicated a break in the planet
population for 2.0$\lesssim$\Rp$\lesssim$2.8
\Rear\ \citep{FRE13,PET13A,PET13B,SIL15}.  (d) Finally, extending this
work to a larger parameter space and for alternative target selection
samples, such as the \kepler\ M dwarf sample where a sharp break at
\Rp$\sim$2.5 \Rear\ is observed \citep{DRE13,BUR15}, the double power
law in \Rp\ is strongly (BIC$>10$) warranted.

Symptomatic of the weak evidence for a broken power law model over the 0.75$\leq$\Rp$\leq$2.5~\Rear\ range, $R_{\rm
  brk}$ is not constrained within the prior \Rp\ limits of the
parameter space.  When $R_{\rm brk}$ is near the lower and upper
\Rp\ limits, $\alpha_{1}$ and $\alpha_{2}$ also become poorly
constrained, respectively.  To provide a more meaningful constraint on
the average power law behavior for \Rp\ in the double power law PLDF
model, we introduce $\alpha_{\rm avg}$, which we set to $\alpha_{\rm
  avg}=\alpha_{1}$ if $R_{\rm brk}\geq R_{\rm mid}$ and $\alpha_{\rm
  avg}=\alpha_{2}$ otherwise, where $R_{\rm mid}$ is the midpoint
between the upper and lower limits of \Rp.  We find $\alpha_{\rm
  avg}=-1.54\pm0.5$ and $\beta=-0.68\pm0.17$ for our baseline result.
We use $\alpha_{\rm avg}$ as a summary statistic for the model
parameters only to enable a simpler comparison of our results to
independent analyses of planet occurrence rates and to approximate the
behavior for the power law \Rp\ dependence if we had used the simpler
single power law model.  The results for a single power law model in
both \Rp\ and \Porb\ are equivalent to the results for the double
power law model ($F_{0}=0.83\pm0.13$, $\alpha=-1.56\pm0.3$, and
$\beta=-0.68\pm0.17$).

In Table~\ref{tab:results}, we provide the parameters of the PLDF that
maximizes the likelihood for the data in our baseline analysis as well
as the median and percentile posterior values for $F_{0}$, $\beta$, and
$\alpha_{\rm avg}$.  Additional statistics for the full five parameter
PLDF can be estimated from the 10,000 posterior MCMC samples in
Table~\ref{tab:baseposterior}.

\begin{deluxetable}{l@{\hspace{4 pt}}c@{\hspace{4 pt}}c@{\hspace{4 pt}}c@{\hspace{4 pt}}c@{\hspace{4 pt}}c@{\hspace{4 pt}}c}
\tablewidth{0pt}
\tablecaption{{\rm PLDF Model Parameter Summary}}
\startdata
\hline
\hline
 & $F_{0}$  & $\alpha_{1}$ & $\alpha_{2}$ & $R_{\rm brk}$ & $\alpha_{\rm avg}$ & $\beta$ \\
\hline
Likelihood Max & 0.73 & 19.68 & -1.78 & 0.94 & ... & -0.65 \\
0.13\% & 0.48 & ... & ... & ... & -3.09 & -1.20 \\
15.9\% & 0.66 & ... & ... & ... & -1.97 & -0.85 \\
50.0\% & 0.77 & ... & ... & ... & -1.54 & -0.68 \\
84.1\% & 0.92 & ... & ... & ... & -1.04 & -0.35 \\
99.9\% & 1.32 & ... & ... & ... &  0.53 & -0.19 \\
Lower Limit & 0.28 & ... & ... & ... & -3.25 & -1.4 \\
Upper Limit & 1.92 & ... & ... & ... & 0.53 & -0.10 \\
\hline
\enddata
\label{tab:results}
\end{deluxetable}

\subsection{Sensitivity Analysis}\label{sec:sensit}

Planet occurrence rate calculations are only as accurate as the
inputs.  The baseline results of Section~\ref{sec:results} represent
our current best set of data that are uniformly applicable to the
\kepler\ observations and planet search results.  The resulting
posterior distribution for the PLDF parameters in the above analysis
only represent their statistical precision and do not capture
potential sources of systematic uncertainties present in the inputs.
To explore the level of systematic errors present in the current
results, we repeat the baseline analysis, but for several scenarios in
which we change a single input.  The following sections describe
results of these sensitivity tests.

\subsubsection{Pipeline Completeness Systematics}\label{sec:sensitcomp}

The pipeline detection efficiency we employ for the baseline analysis
is calibrated with transit injection and recovery tests \citep{CHR15},
but it represents the pipeline response averaged over a wider range of
\kepler\ targets than the limited GK dwarf sample of this study.  In
addition, \citet{CHR15} analyzed a shorter (four \kepler\ quarter)
subset of the entire Q1-Q16 data.  It is expected that the pipeline
completeness primarily depends upon the MES and number of transits,
thus the results from the shorter four quarter analysis are applicable
to the sixteen quarter run.  However, star-by-star deviations are
expected, and until we perform larger injection studies it is prudent
to investigate the sensitivity of the occurrence rates to this
potential source of uncertainty.  We consider an optimistic and
pessimistic detection efficiency relative to the baseline result.  For
the optimistic detection efficiency, we assume the ideal theoretical
expected performance of TPS given by Equation~(\ref{eq:tpsthy}).  For
the pessimistic detection efficiency we assume the result from
\citet{FRE13}, where they find a linear detection efficiency over the
range 6$<$MES$<$16 provides the best match to the SNR distribution of
the Q1-Q6 \kepler\ planet candidate sample \citep{BAT13}.  The
detection efficiency of the Q1-Q6 \kepler\ pipeline was never measured
directly using Monte-Carlo transit recovery tests.  Thus, we cannot
determine the accuracy of the \citet{FRE13} detection efficiency
relative to the Q1-Q6 \kepler\ pipeline run.  However, having measured
the detection efficiency for the Q1-Q16 pipeline run \citep{CHR15},
the \citet{FRE13} detection efficiency is overly pessimistic for the pipeline completeness of the Q1-Q16 pipeline run.

\begin{figure}
\includegraphics[trim=0.25in 0.05in 0.35in 0.3in,scale=0.5,clip=true]{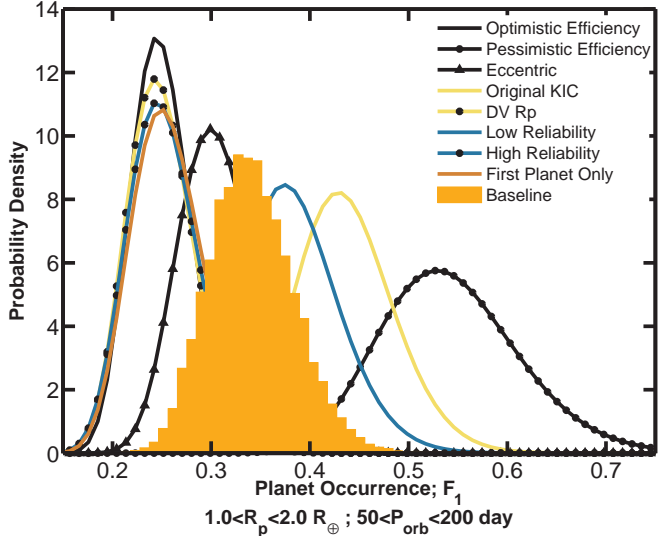}
\caption{Posterior distribution for the integrated planet occurrence rate over the 1.0$<$\Rp$<$2.0 \Rear\ and 50$<$\Porb$<$200~days parameter space, $F_{1}$.  Changes from the baseline inputs (filled orange histogram) systematically impact the derived occurrence rate beyond the statistical uncertainty.  We discuss in Section~\ref{sec:sensit} alternative inputs: optimistic detection efficiency (black line), pessimistic detection efficiency (black with circles line), assuming e=0.4 for all orbits (black with triangles line), original KIC stellar parameters (yellow line), alternative DV \Rp\ (yellow with circles line), low reliability planet candidate sample (blue line), high reliability planet candidate sample (blue with circles line), and assuming a single planet search (red line).\label{fig:syscomp}}
\end{figure}

Overall, an overly optimistic detection efficiency
reduces the planet occurrence rate and a pessimistic detection
efficiency increases the planet occurrence rates.  We show in
Figure~\ref{fig:syscomp} the posterior integrated planet occurrence
rate for the baseline result (orange histogram) compared to the case
of an optimistic (black line) and the pessimistic (black with circles
line) detection efficiency alternatives.  For this comparison we
narrow the parameter space of integration (1.0$<$\Rp$<$2.0 \Rear\ and
50$<$\Porb$<$200~days) in order to limit the comparison to a region
of parameter space with better statistics and to facilitate comparison
with \kepler\ occurrence rates from previous studies.  We symbolize
this narrower parameter space planet occurrence rate as $F_{1}$.  For
clarity of display in Figure~\ref{fig:syscomp}, the optimistic and
pessimistic occurrence rate posteriors are shown by a log-normal fit
to the posterior rather than the full posterior detail in a histogram
format.  The pessimistic detection efficiency has a $>$3-$\sigma$
larger occurrence rate than the baseline result and the optimistic
detection efficiency is 2.5-$\sigma$ lower than the baseline result.
This initial test demonstrates that systematic effects can be larger
than the random uncertainties.

\begin{figure}
\includegraphics[trim=0.25in 0.05in 0.35in 0.3in,scale=0.5,clip=true]{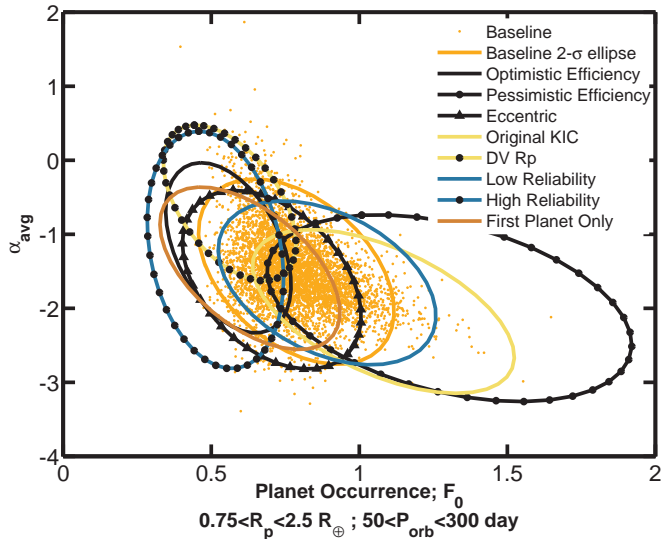}
\caption{Samples from the posterior distribution of $F_{0}$ as a function $\alpha_{\rm avg}$ for the baseline results (orange points) along with an approximate 2-$\sigma$ error ellipse for the baseline results (orange ellipse).  Also shown are 2-$\sigma$ error ellipses for the alternative inputs with the same line types as in Figure~\ref{fig:syscomp}.\label{fig:sensit_alpha}}
\end{figure}

\begin{figure}
\includegraphics[trim=0.25in 0.05in 0.35in 0.3in,scale=0.5,clip=true]{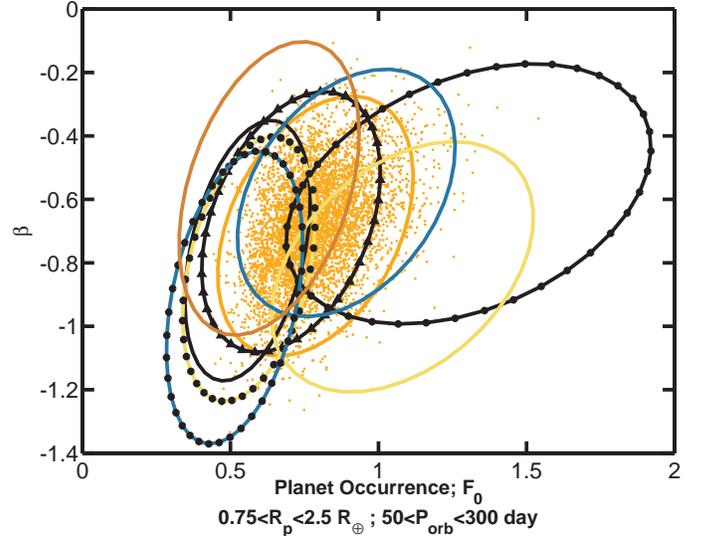}
\caption{Same as Figure~\ref{fig:sensit_alpha}, but showing the posterior distribution of $F_{0}$ as a function $\beta$.\label{fig:sensit_beta}}
\end{figure}

The alternative inputs also influence the other `shape' parameters of
the PLDF.  We show samples from the posterior distribution of
$\alpha_{\rm avg}$ (Figure~\ref{fig:sensit_alpha}) and $\beta$
(Figure~\ref{fig:sensit_beta}) as a function of $F_{0}$ for the baseline
(orange points) occurrence rate parameter estimates.  As an
approximation to the joint 2-$\sigma$ posterior distribution we model
the posterior as a multi-normal distribution with major and minor axes
along the eigenvectors determined from the posterior samples (orange
ellipse).  For comparison, we show the optimistic (black ellipse) and
pessimistic (black with circles ellipse) detection efficiency solutions by the
2-$\sigma$ ellipse model for the shape parameters.  The systematic
variations of $\alpha_{\rm avg}$ and $\beta$ are correlated with $F_{0}$.

\subsubsection{Orbital Eccentricity}\label{sec:sensitecc}

In the baseline result we have assumed circular orbits when
constructing the model for pipeline completeness.  However, radial
velocity studies have revealed that eccentric orbits are common for
\Porb$>$10 days \citep{BUT06}. A nonzero eccentricity results in higher
probability to transit, but a shorter transit duration degrades the
transit SNR \citep{BUR08}.  \citet{BUR08} shows that yields
from a transit survey could be up to 25\% higher using the observed
distribution of radial velocity planets.  We investigate a limiting
case of assuming all planets have e=0.4.  The e=0.4 case results in an
11\% (1-$\sigma$) lower planet occurrence rate (black with triangles line in Figure~\ref{fig:syscomp}).  Thus, for this parameter space the
systematic effect due to orbital eccentricity is comparable to the
statistical errors.  The impact of eccentricity on $\alpha_{\rm avg}$
and $\beta$ is also modest.

\subsubsection{Stellar Parameter Systematics}\label{sec:sensitstar}

Stellar parameter estimates of \kepler\ targets are subject to
systematic uncertainties
\citep{BRO11,MAN12,MUI12,PIN12,DRE13,EVE13,GAI13} and
multiplicity/blend effects \citep{CAR14,LIL14,CIA15}.  Also, the
stellar parameter catalog of \citet{HUB14} relies upon a heterogeneous
compilation of input sources and still has some limitations (see
their Section~8 for a discussion).  As a proxy for constraining the
impact on occurrence rate studies due to stellar parameter
systematics, we repeat the analysis but adopt stellar parameters from
the original KIC \citep{BRO11}.  Using the original KIC is also
germane since it was employed for previous work on planet
occurrence rates with \kepler\ \citep{HOW12,FRE13}.

We redo the GK dwarf target selection resulting in 102,186 targets
that meet the selection criteria.  There are 83,724 targets (91.4\%)
in common with the baseline GK dwarf sample discussed in
Section~\ref{sec:stars}.  Table~\ref{tab:kicused} provides a binary
flag to indicate that the \kepler\ target was included in the stellar
sample based upon the original KIC stellar parameters.  In the
original KIC GK dwarf sample there are 177 planet candidates that have
122 planet candidates (78.2\%) in common with the baseline planet
candidate sample discussed in Section~\ref{sec:planets}.
Table~\ref{tab:koiused} has a binary flag indicating the planet
candidates selected for this original KIC stellar sample.  We adjust
the derived \Rp\ of the planet candidate sample by linearly scaling
\Rp\ by the ratio in \Rstar\ between the baseline and the original KIC
values.

The net impact of the alternative stellar parameters results in $\sim
2\sigma$ higher occurrence rates (yellow line in
Figure~\ref{fig:syscomp}).  The change in $\alpha_{\rm avg}$ is larger
than for $\beta$ (yellow ellipse in
Figures~\ref{fig:sensit_alpha}~and~\ref{fig:sensit_beta}).

\subsubsection{Planet Candidate Parameters}~\label{sec:sensitplanets}

Planet radii are not a direct observable, and they must be derived
through parameter fits to light curves with various assumptions as to
the stellar parameters, limb darkening coefficients, flux time series
detrending, treatment of instrumental effects, orbital eccentricity,
and third light contamination to name a few
\citep{MAN02,SEA03,GIM06,HOL06,BUR07,TOR08,SOU11,KIP14,ROW15,MULL15,CIA15}.
In our current analysis, we treat \Rp\ as perfectly known without
uncertainty.  Recent work has pointed out the non-negligible bias in
deriving planet occurrence rates without taking into account the full
error distribution of \Rp\ \citep{MOR14B,FOR14,SIL15,DRE15}.  Detailed
planet parameter posterior estimates have only recently become
available for a majority of the \kepler\ planet candidate sample
\citep{ROW15B}, thus we defer occurrence rate analysis using a full
posterior distribution of planet radii for future work.

We repeat the occurrence rate calculation using the
alternative \Rp\ estimates provided by the Data Validation (DV) module
of the \kepler\ pipeline \citep{WU10}.  The most important differences
between the DV analysis and the baseline planet parameters from
\citet{MULL15} and \citet{ROW14} are the independent methods of
detrending the flux time series data and DV use of $\chi^{2}$
minimization instead of a MCMC analysis.  Both analyses assume the
same stellar parameters, fixed limb darkening coefficients, zero
eccentricity, and begin with the pre-search data conditioning time
series \citep{STU12,SMI12}.  \citet{MULL15} find that the radii ratios,
\Rp/\Rstar, from the MCMC analysis are $\sim$7\%
smaller than from the analysis in DV.
The typically larger \Rp\ from DV
results in $\sim 2.2\sigma$ lower occurrence rates (yellow with
circles line in Figure~\ref{fig:syscomp}).  The change in $\alpha_{\rm
  avg}$ is larger than for $\beta$ (yellow with circles ellipse in
Figures~\ref{fig:sensit_alpha}~and~\ref{fig:sensit_beta}).

\subsubsection{Planet Candidate Sample}

Characterizing a detection by the \kepler\ pipeline as a bona fide
member of the \kepler\ planet candidate sample has increasingly relied
upon an automated classification procedure
\citep{MCC14,ROW15,MULL15,COU15,ROW15}.  However, the accuracy,
efficacy, and impact on deriving planet occurrence rates due to the
automated classification and remaining manual vetting decision steps
have not been fully quantified.  The vetting process has its own false
negative alarm rate outside of the pipeline completeness, that
currently we do not account for.  In addition, the decision process
for both the automated and manual decision methods becomes
increasingly less definitive towards low SNR \citep[see the discussion
  of the current planet sample limitations in Sections~7~and~9.1
  of][]{MULL15}.  The planet candidate catalog of \citet{MULL15} takes
an `innocent until proven guilty' approach to deal with the
indeterminant diagnostics in the low SNR regime.  The instrumental
aperture contamination and crosstalk also become increasingly
difficult to identify at low SNR \citep{COU14}.  We constrain the
potential systematics in deriving planet occurrence rates due to
uncertainty in the planet candidate sample classification process by
considering two alternative planet samples.

First, we include a population of twelve `lower reliability' KOIs with
a false positive disposition in the 50$\leq$\Porb$\leq$300 days and
\Rp$\leq$2.5 \Rear\ parameter space under investigation (see
Table~\ref{tab:lowreliable}).  This sample of `lower reliability' KOIs
were characterized as planet candidates for all the vetting procedures
described in \citet{MULL15} except one.  These KOIs are false
positives because they failed to maintain an SNR$\geq7.1$ in the
independent detrending employed for the MCMC planet parameter
estimates \citep{ROW14}.  Prior to the MCMC evaluation, a trial
$\chi^{2}$ minimization provides a parameter initialization.  These
lower reliability KOIs failed to yield SNR$\geq7.1$ in this trial fit
and were therefore demoted from a PLANET CANDIDATE to a FALSE POSITIVE disposition.  Requiring an independent recovery of a
potential detection is a valuable criteria for a planet candidate, but
it largely impacts our lowest SNR detections and we have not fully
quantified the false negative rate of this independent recovery test.
In lieu of a more detailed investigation, it provides a useful
limiting test case sample to constrain the potential breakdown of the
vetting metrics at the lowest SNR of the planet candidate sample.
Including a lower reliability KOI sample in the planet candidate list,
results in $\sim 1\sigma$ higher occurrence rates (blue line in
Figure~\ref{fig:syscomp}) and modest changes in $\alpha_{\rm avg}$ and
$\beta$ (blue ellipse in
Figures~\ref{fig:sensit_alpha}~and~\ref{fig:sensit_beta}).

\begin{deluxetable}{c}
\tablewidth{0pt}
\tablecaption{{\rm Low Reliability KOI False Positive Sample}}
\startdata
\hline
\hline
KOI number \\
\hline
4954.01 \\
5043.01 \\
5081.01 \\
5102.01 \\
5123.01 \\
5177.01 \\
5198.01 \\
5210.01 \\
5257.01 \\
5309.01 \\
5325.01 \\
5405.01 \\
\hline
\enddata
\label{tab:lowreliable}
\end{deluxetable}

Second, we cull the baseline KOI planet candidate sample to the most reliable
detections by requiring KOIs to have been detected in at least one
other pipeline run.  Each run of the \kepler\ pipeline is independent
and has different amounts and versions of the data.  To
remain in the `high reliability' planet candidate sample, we require a
KOI to be represented as a TCE in either the Q1-Q12 pipeline run
\citep{TEN13,ROW15}, the Q1-Q17 pipeline run \citep{SEA15}, or a
testing/development run using Q1-Q17 data with a near-final
\kepler\ pipeline code version.  This requirement removed 26 KOI
planet candidates in the 50$\leq$\Porb$\leq$300 days and \Rp$\leq$2.5
\Rear\ parameter space under investigation.  Table~\ref{tab:koiused}
contains a binary flag indicating the planet candidates belonging to
this high reliability planet sample.  Adopting a higher reliability
KOI sample results in $\sim 2.2\sigma$ lower occurrence rates (blue with circles line in Figure~\ref{fig:syscomp}).  The change in $\alpha_{\rm
  avg}$ and $\beta$ (blue with circles ellipse in
Figures~\ref{fig:sensit_alpha}~and~\ref{fig:sensit_beta}) are
consistent with preferentially removing the lower SNR KOIs which
typically reside at smaller radii and longer orbital periods.

The final systematic we investigate is the impact of limiting the
search to a single planet per target, effectively ignoring the
multiple planet search in the \kepler\ pipeline.  We provide this
result in order to more directly compare independent analyses of
the \kepler\ data that do not search for multiple planets
\citep{PET13B}.  \citet{PET13B} estimate that their occurrence rates
would be $\sim$25\% higher by including multiple planet systems in their
study.  We concur with their estimate by finding a 25\% ($\sim
2.2\sigma$) lower occurrence rate by only including the lowest
numbered KOI (typically the highest SNR) of a system (red line in
Figure~\ref{fig:syscomp}).  The change in $\alpha_{\rm avg}$ is
negligible and $\beta$ prefers a more gradual decrease in planet
occurrence with \Porb\ despite the lower overall occurrence rate
normalization (red ellipse in
Figures~\ref{fig:sensit_alpha}~and~\ref{fig:sensit_beta}).

\subsubsection{Systematic Error Summary}

The previous sections show that individual systematic effects can
reach 2$\sigma$ biases in the occurrence rates, where $\sigma$ is
determined from statistical errors alone.  Unfortunately, multiple
systematic effects can add coherently rather than quadratically (see
Section~\ref{sec:disc}).  To provide a more realistic uncertainty in
the context of all these systematic uncertainties, we express the
uncertainties on the occurrence rate parameters as an acceptable range.
We adopt the lower and upper limit of the acceptable range as the
2$\sigma$ lower and upper limit for the largest systematic effect
calculated in the previous sections.  Based upon the results in this
section, we find that the planet occurrence rate for the
1.0$<$\Rp$<$2.0~\Rear\ and 50$<$\Porb$<$200~days parameter space to
have a best value from the baseline calculation of $F_{1}=0.34$ with an
acceptable range of 0.19$\leq F_{1} \leq$0.7.  We provide acceptable
ranges for the PLDF model parameters in Table~\ref{tab:results}.

\section{Discussion}\label{sec:disc}

In this section, we compare our PLDF to previous work on the
\kepler\ target sample that included analysis of the G dwarf targets
using at least twelve quarters of \kepler\ data.  We generally find higher
occurrence rates, no evidence for a break at \Rp$\lesssim$2.5 \Rear,
increasing planet occurrence rates towards \Rp=1.0 \Rear, slightly
shallower drop-off of occurrence rates towards longer \Porb, and
larger uncertainty on occurrence rates driven by systematic effects.

\begin{figure}
\includegraphics[trim=0.25in 0.05in 0.35in 0.3in,scale=0.5,clip=true]{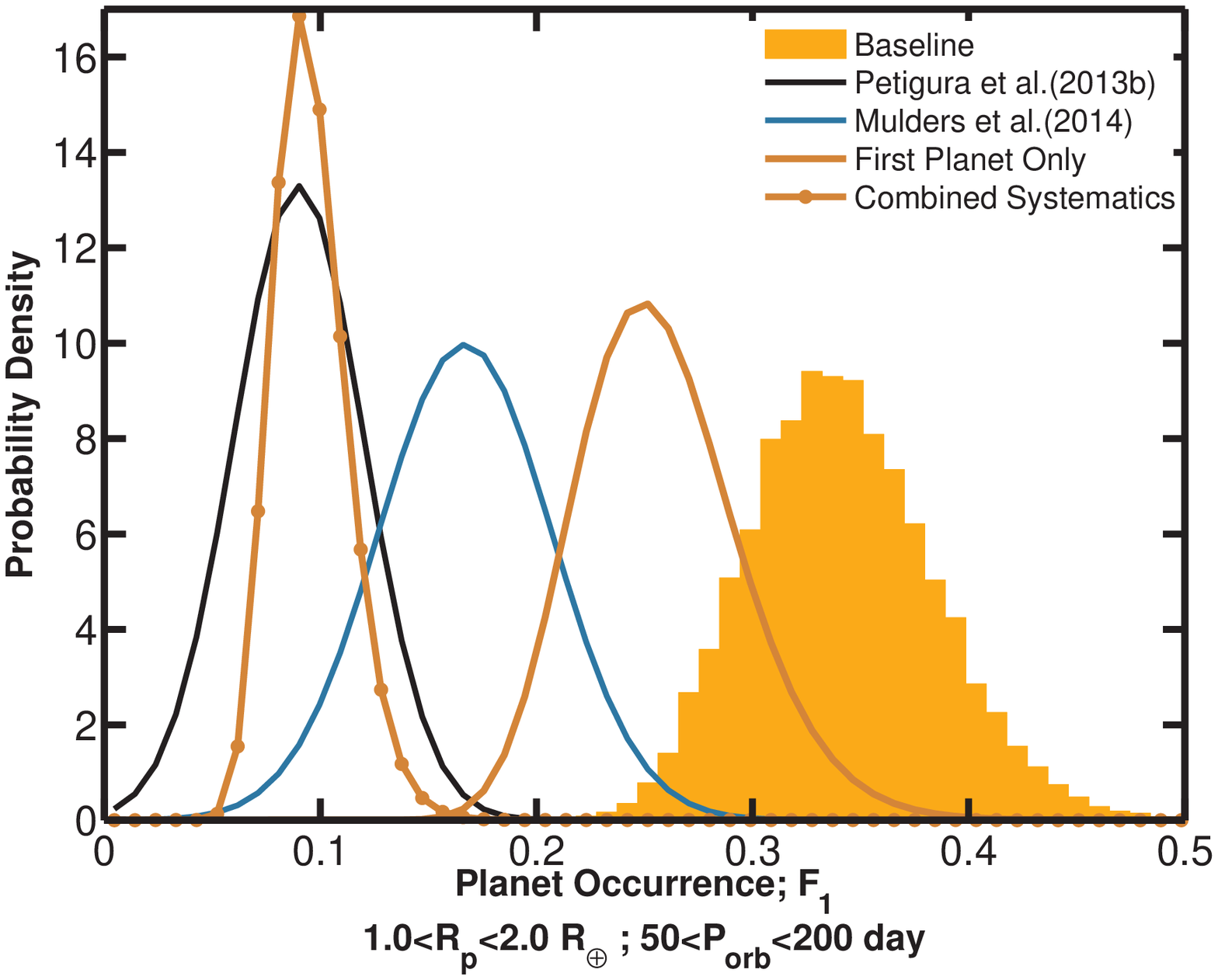}
\caption{Comparison of the integrated posterior distribution from our baseline PLDF over the 1.0$<$\Rp$<$2.0 \Rear\ and 50$<$\Porb$<$200~days parameter space (orange filled histogram) to previous results over the same parameter space from \citet{PET13B} (black curve) and \citet{MUL15} (blue curve).  The posterior distribution from the other works are approximated as Gaussians with their central and standard deviation parameters as published.  We also show our single planet search results (red line) and an extreme scenario where the four leading systematics resulting in lower occurrence rates are combined (red with circles line, see Section~\ref{sec:disc}).\label{fig:compoccurr}}
\end{figure}

As a primary source for comparison, we compare to the independent
pipeline analysis on planet occurrence rates by \citet{PET13B}.  We
compare the integrated planet occurrence rate over the
1.0$<$\Rp$<$2.0~\Rear\ and 50$<$\Porb$<$200~days range, $F_{1}$, in
Figure~\ref{fig:compoccurr}.  We approximate the result from
\citet{PET13B} ($F_{1}$=9$\pm$3\% occurrence rate) as a Gaussian
(black line) with value and uncertainty as published from their
Figure~2.  The posterior distribution of our baseline result (orange
histogram) demonstrates a significant difference from the occurrence
rate of \citet{PET13B}.  For consistency with the TERRA pipeline
(which does not search for multiple planets), we show our alternative
occurrence rate after keeping only the highest SNR planet candidate
for a target (red line) in Figure~\ref{fig:compoccurr}.  The `first
planet only' occurrence rate does not fully remove the difference.
In our analysis we explored numerous alternative inputs (see
Section~\ref{sec:sensit}). Even when assuming a wide variety of systematics,
we have a difficult time reconciling our results with \citet{PET13B}.

It is possible that several sources of systematics add coherently to
reconcile the results.  For instance, we can reproduce the occurrence
rate of \citet{PET13B} (black line in Figure~\ref{fig:compoccurr}) by
combining together four of the systematic effects resulting in lower
occurrence rates: single planet search only, alternative DV \Rp,
highest reliability KOIs, and optimistic detection efficiency.  The
resulting occurrence rate (red with circles line in
Figure~\ref{fig:compoccurr}) $F_{1}=0.09$ is less than our lower limit
of an acceptable range $F_{1}\geq0.19$.  Further work is needed to
understand the differences between our results and the results of
\citet{PET13B}.  Some possibilities including increasing the number of
injection and recovery trials, characterizing the impact of the flux
time series detrending on planet recovery, investigating systematic
differences between the stellar parameter estimates of the planet
candidate hosts and non planet candidate hosts, and better
characterization of the biases that may be present when using the
binned occurrence rate methodology \citep{MOR14B}.  {\cjb For
  instance, \citet{FOR14} note a bias in the binned occurrence rate
  methodology is present if the completeness function is evaluated at
  exactly the location of the planet parameters rather than being
  averaged over the entire bin of analysis.}

One characteristic result of \citet{PET13B} is a plateau to declining
occurrence rates in the mini-Neptune to terrestrial planet
regime.  To enable comparison of the results from our parametric PLDF
model, we derive an approximate PLDF consistent with the occurrence
rate from Figure~3 of \citet{PET13B}, marginalized over
5$<$\Porb$<$100~days.  We determine a PLDF with two free parameters,
$g_{\rm p}=F_{\rm p}R_{\rm p}^{\alpha_{\rm p}}P_{\rm orb}^{-1}$, using
the 1$\leq$\Rp$\leq$1.4~\Rear\ and 2$\leq$\Rp$\leq$2.8~\Rear\ bins
from Figure~3 of \citet{PET13B}, yielding $\alpha_{\rm p}=-0.3677$ and
$F_{\rm p}=0.103$.  The approximating PLDF yields an occurrence rate
of 14.9\% for the 1.4$\leq$\Rp$\leq$2.0~\Rear\ bin compared to the
published value of 14.2$\pm$1.0\% from \citet{PET13B}.  As a
  further check, the approximating PLDF yields a $F_{1}$=12.5\%
  occurrence rate compared to $F_{1}$=9$\pm$3\% for the published
  1.0$<$\Rp$<$2.0 \Rear\ and 50$<$\Porb$<$200~days rate of
  \citet{PET13B}.  The derived $\alpha_{\rm p}$ is $\sim$2-$\sigma$
(statistical uncertainty alone) shallower than our $\alpha_{\rm
  avg}=-1.5$ power law dependence of the occurrence rates on \Rp.
However, $\alpha_{\rm p}$ is consistent with our results for
$\alpha_{\rm avg}$ if one considers the full systematic range
-3.25$\leq\alpha_{\rm avg}\leq$0.53 allowed from the sensitivity
analysis of Section~\ref{sec:sensit}.  A similar comparison applies to
\Porb\ dependence of the PLDF for the $\beta$ parameter.  Our
$\beta=-0.68$ is $\sim$2-$\sigma$ (statistical uncertainty alone)
shallower than the $\beta_{\rm p}=-1$ dependence qualitatively stated
in \citet{PET13B}.  However within the full range allowed,
-1.4$\leq\beta\leq$-0.1, the two values agree.  We show the
approximating PLDF (dash dot line) for comparison to our result in
Figure~\ref{fig:smallrp_rp_underlying}.\footnote{The rising slope
  toward smaller \Rp\ of the approximating PLDF model from the
  \citet{PET13B} results is visually inconsistent with the decreasing
  occurrence rate shown in Figure~3 of \citet{PET13B}, but the visual
  inconsistency arises due to our adoption of linear bin widths for
  this study and the adoption of logarithmic bin widths of
  \citet{PET13B}.  Thus, $\alpha=0.0$ corresponds to a flat occurrence
  rate in the linear bin widths of this study and $\alpha=-1.0$ would
  correspond to a flat occurrence if we were to adopt logarithmic bin
  widths.}

If we use our baseline inputs to fit the broken powerlaw in \Rp\ PLDF
over a larger, 0.75$<$\Rp$<$5.0 \Rear, parameter space, we do find
decisive evidence, BIC$>$10, for the broken power law model over a
single powerlaw in \Rp\ PLDF.  The derived $R_{\rm
  brk}=3.3^{+0.2}_{-0.4}$ \Rear, with a $\alpha_{1}=-1.72\pm0.3$ power
law dependence for \Rp$<R_{\rm brk}$ and $\alpha_{2}=-6.6\pm1.7$.  The
planet occurrence rate derived from this study is consistent with a
power law break, but we find that it qualitatively occurs at a larger
radius than the study of \citet{PET13A} ($R_{\rm brk}\sim 2.5$), {\cjb but is consistent with the qualitatively stated break at $R_{\rm brk}\sim 3$ of \citet{DON13}.}

We also compare to the integrated occurrence rate from the \kepler\ G
dwarf sample of \citet{MUL15} (blue line) in
Figure~\ref{fig:compoccurr}.  To estimate a value from Table~7 of
\citet{MUL15}, the 150$<$\Porb$<$250 bin was weighted by 0.56 assuming
a \Porb$^{-1}$ PLDF dependence across the bin.  This occurrence rate
is in between the results of \citet{PET13B} and this study, and has
uncertainty overlap with both studies especially when considering the
systematic sources of error.  {\cjb We find a very similar result
  between \citet{MUL15} and \citet{DON13} for the occurrence rate in
  this parameter space.}  \citet{SIL15} find results comparable to
\citet{PET13B}.

\section{Extrapolation to Longer Periods}\label{sec:extrap}

In this section, we compare the observed Q1-Q16 planet candidate
sample at longer periods (300$<$\Porb$<$700~days) to the predicted planet
candidate yield deduced by extrapolating the PLDF model with
parameters determined from the shorter period (50$<$\Porb$<$300~days) parameter
space.  In our baseline study, we purposely avoided the longer period regime
because the planet candidate sample with three transit events and low MES has
the potential for a substantially higher false alarm rate \citep[see
  the discussion in Section~9.1 of][]{MULL15}.  In previous planet
candidate samples, the false alarm rate was minimal since a
KOI detection from an earlier pipeline run could be compared to a
later pipeline run with substantially more data available.  With the
ending of the \kepler\ primary mission, further data beyond Q1-Q17 is
not available to verify our lowest MES detections.

\begin{figure}
\includegraphics[trim=0.25in 0.05in 0.35in 0.3in,scale=0.5,clip=true]{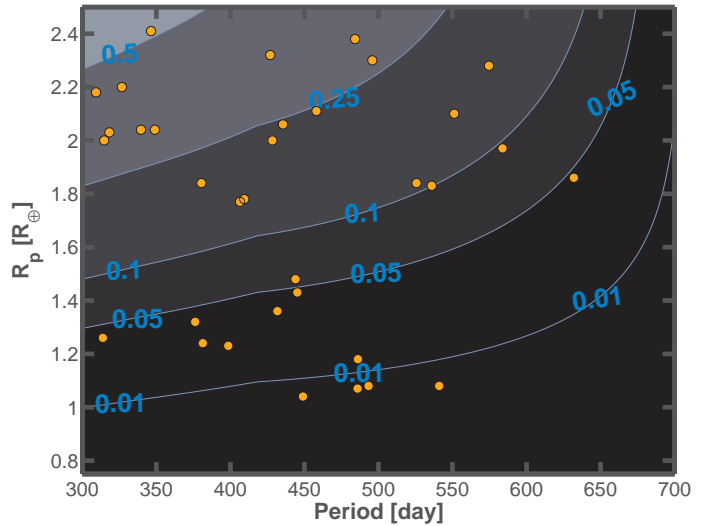}
\caption{Average pipeline completeness contours for the GK dwarf sample toward longer, 300$<$\Porb$<$700~days, along with the Q1-Q16 \kepler\ planet candidate sample (orange points).\label{fig:longpavgdet}}
\end{figure}

\begin{figure}
\includegraphics[trim=0.05in 0.05in 0.35in 0.3in,scale=0.5,clip=true]{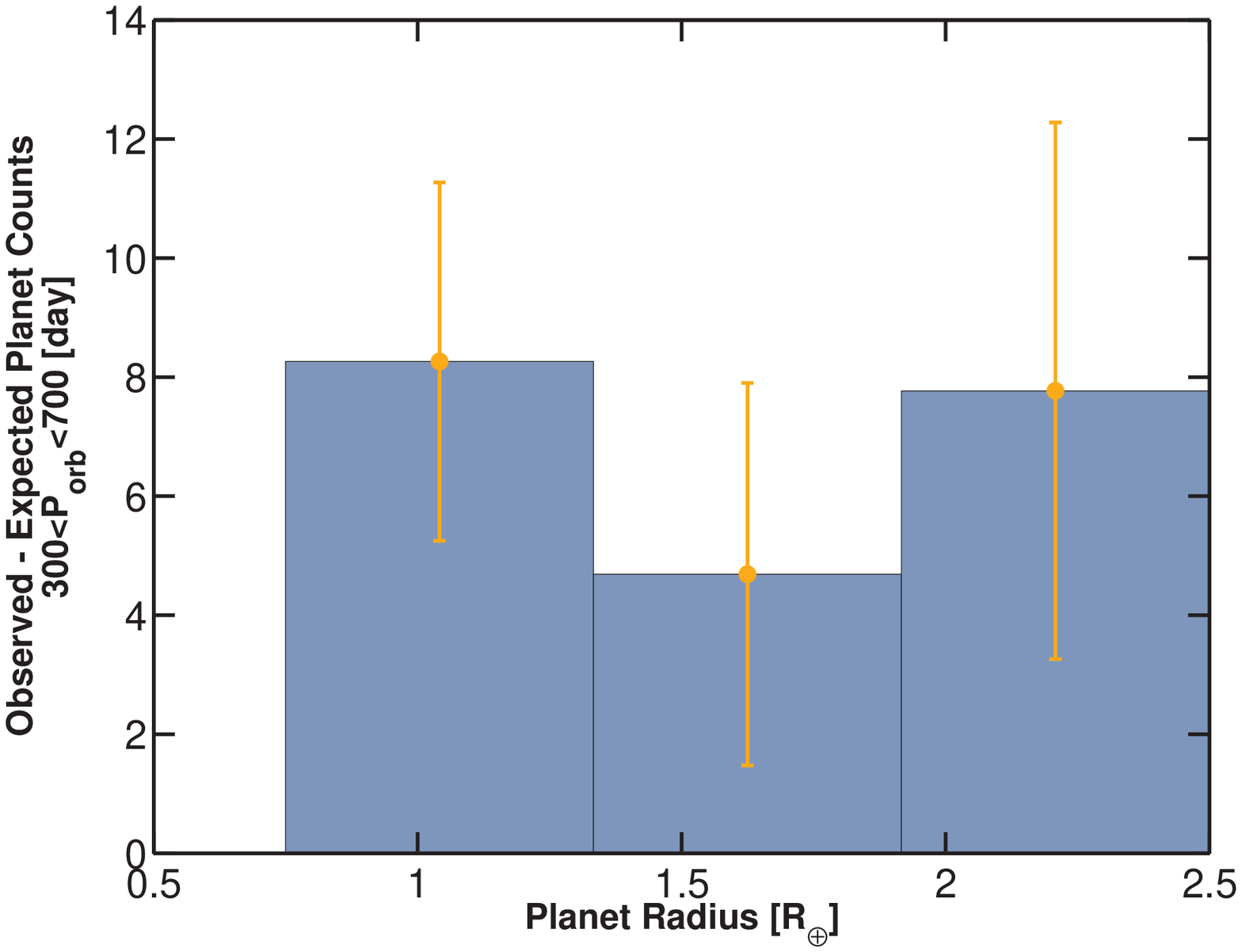}

\includegraphics[trim=0.05in 0.05in 0.35in 0.3in,scale=0.5,clip=true]{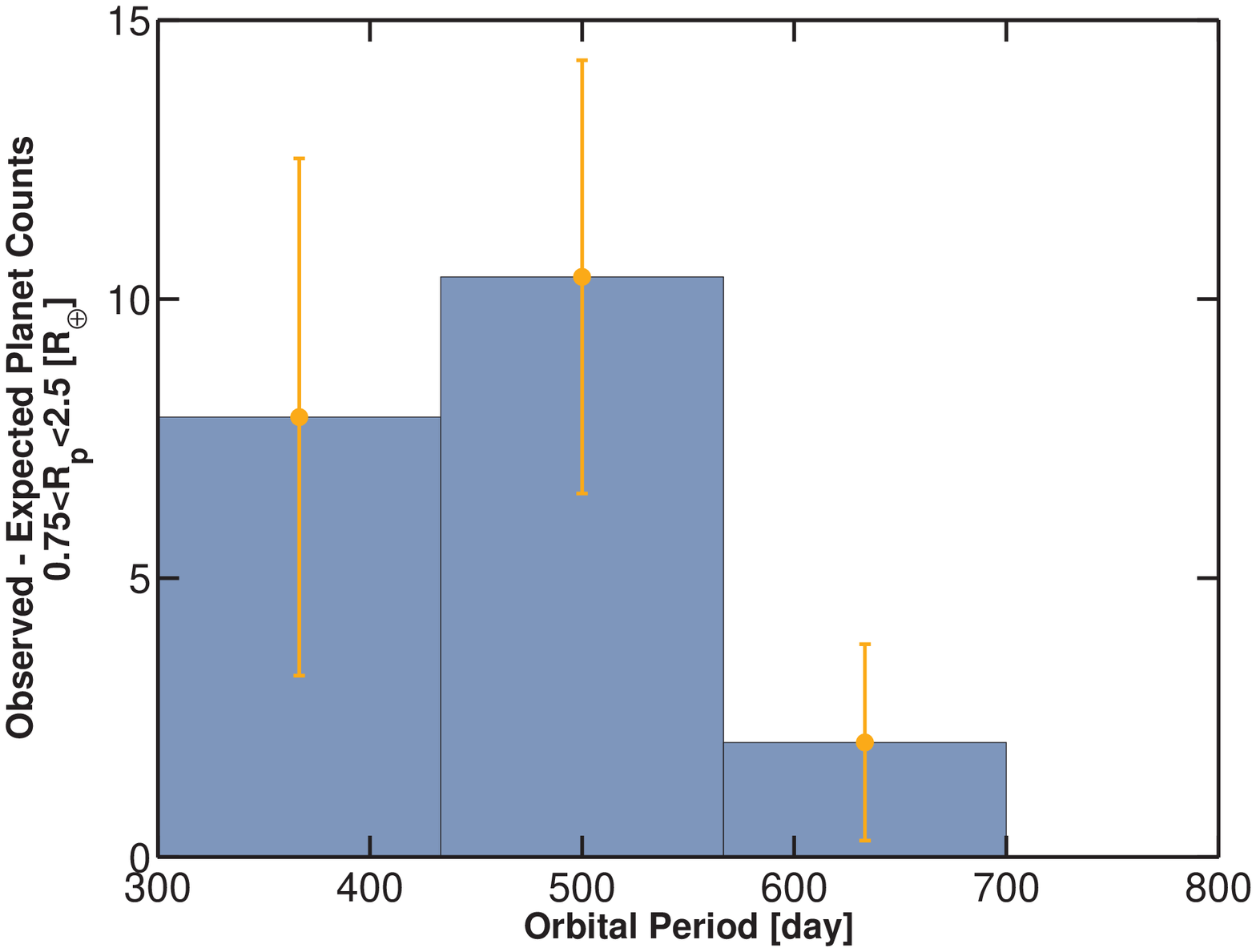}
\caption{Top: marginalized over periods of 300$<$\Porb$<$700~days observed \kepler\ planet candidate counts minus the predicted planet candidate counts obtained by extrapolating our planet occurrence rate results from the shorter 50$<$\Porb$<$300~days analysis of Section~\ref{sec:results}.  Bottom: same as top, but marginalized over planet radius 0.75$<$\Rp$<$2.5 \Rear.\label{fig:predict}}
\end{figure}

Figure~\ref{fig:longpavgdet} shows the average pipeline completeness
contours toward longer \Porb\ for the GK dwarf sample of this study
along with the \kepler\ planet candidate sample in this regime.  Using
this long period pipeline completeness model and the shorter period
PLDF model, we predict the expected planet candidate yield for
\kepler.  The top panel of Figure~\ref{fig:predict} shows the
difference between observed and predicted planet candidate counts
marginalized over 300$<$\Porb$<$700~days.  There is a statistically
significant overabundance of planet candidates toward longer periods
than predicted from the baseline PLDF derived at the shorter orbital
periods.  The largest discrepancy is for the smallest \Rp\ bin under
consideration in Figure~\ref{fig:predict}.  The bottom panel of
Figure~\ref{fig:predict} shows the observed minus predicted planet
candidate counts as a function of \Porb\ after marginalizing over
0.75$<$\Rp$<$2.5 \Rear.  The most significant overabundance is in the
middle \Porb\ bin.  The largest contributor to the overabundance are
the cluster of five planet candidates around \Rp$\sim$1.1 \Rear\ and
\Porb$\sim$500~days that fall along the (0.01) average pipeline
completeness contour level.

The significant overabundance of planet candidates implies that
extrapolations of our PLDF from the 50$<$\Porb$<$300~days range may
underestimate the planet occurrence rates toward longer periods.
However, at this time we cannot distinguish between a higher
occurrence of planets toward long periods in the \kepler\ GK dwarf
planets, a larger false alarm contribution among the lowest MES planet
candidates, or systematic bias in our simplified pipeline completeness
model.  We are investigating flux time series inversion and
permutation tests along with a bootstrap noise characterization test
\citep{SEA15} in order to calibrate the false alarm rate in the
\kepler\ planet candidate sample.

\section{Terrestrial Planet Occurrence Rate For Venus Orbital Periods}\label{sec:venus}

Earth's sister planet, Venus, has an orbital period within the
\Porb\ range of the baseline analysis.  Thus, in this section we
present results for the occurrence rate of terrestrial planets
corresponding to the \Porb$\sim$0.6~yr of Venus.  We define
$\zeta_{0.6}$ as the 0.6~yr terrestrial planet occurrence rate, which
we take to be within 20\% of \Rp=1~\Rear\ and 20\% of
\Porb$_{\venus}$.  The integral range of 20\% is within the
expectations for the regime of rocky terrestrial planets
\citep{ROG15,WOL14}.  Since \Porb\ is a direct observable, providing
occurrence rates in terms of \Porb\, such as $\zeta$, has advantages
over providing occurrence rates in terms of stellar insolation flux,
such as the Venus zone ($\eta_{\venus}$) concept of \citet{KAN14} or
the HZ ($\eta_{\oplus}$) concept \citep{KAS93,SEL07,ZSO13,KOP13}.
Stellar insolation flux is an indirectly measured quantity and
$\eta_{\venus}$ and $\eta_{\oplus}$ depend upon uncertain theoretical
models for terrestrial planet atmospheric evolution.  Providing
occurrence rates in terms of \Porb\ facilitates comparison with future
\kepler\ occurrence rate studies and is readily compared to
theoretical terrestrial planet formation models.

\begin{figure}
\includegraphics[trim=0.25in 0.0in 0.35in 0.05in,scale=0.5,clip=true]{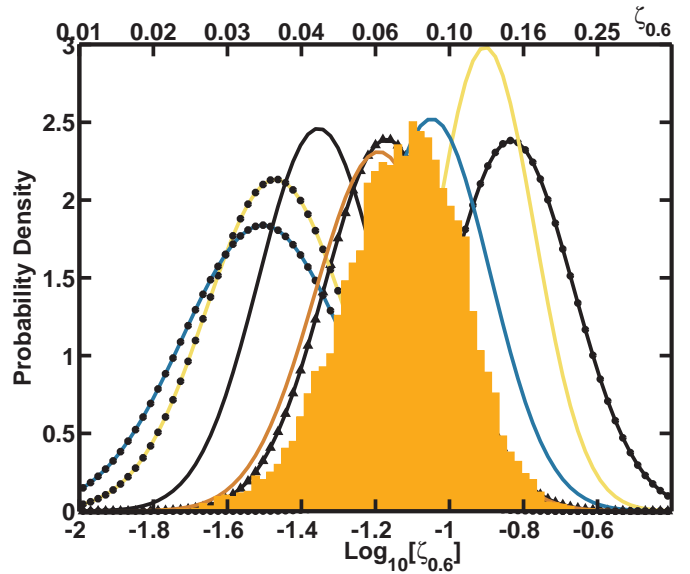}
\caption{Distribution for the 0.6~yr terrestrial planet occurrence rate, $\zeta_{0.6}$, integrated within 20\% of \Rp=1~\Rear\ and \Porb$_{\venus}$, using the baseline analysis (filled orange histogram).  Solid lines represent results using alternative inputs with the same line types as in Figure~\ref{fig:syscomp}.\label{fig:venus}}
\end{figure}

We defer the additional complications in calculating $\eta_{\venus}$
and $\eta_{\oplus}$ to future work.  Despite the complications, for
G~dwarfs, $\zeta_{0.6}$ is a subset of the full $\eta_{\venus}$
parameter space, thus $\zeta_{0.6}$ places a valuable lower limit on
$\eta_{\venus}$ for G~dwarfs.  For the K~dwarfs, \Porb$_{\venus}$
corresponds to the Sun-Earth insolation flux.  Thus,
$\zeta_{0.6}$ is a lower limit on the K dwarf $\eta_{\oplus}$.  We
find $\zeta_{0.6}=0.075$ with an acceptable range of
0.013$\leq\zeta_{0.6}\leq$0.30, and show the baseline and
systematic posterior distributions for $\zeta_{0.6}$ in
Figure~\ref{fig:venus}.

\section{Terrestrial Planet Occurrence Rate For One Year Orbital Periods}\label{sec:earth}

\subsection{Extrapolating to One Year Orbital Period}

The longer, 300$<$\Porb$<$700~days parameter space roughly coincides with the theoretical HZ for the G~dwarf targets,
which is a preferred location in a planetary system for a stable water
bearing planet \citep{KAS93}.  In Section~\ref{sec:extrap}, we
demonstrated that determining the planet occurrence rate in the
300$<$\Porb$<$700~days range directly from \kepler\ data is at a
premature stage due to significant false alarm contamination.  In this
section, we extrapolate our PLDF parametric model in order to
calculate two occurrence rate parameters that can be used as a
baseline for comparison to future work on refining HZ occurrence rates.

\begin{figure}
\includegraphics[trim=0.25in 0.0in 0.35in 0.05in,scale=0.5,clip=true]{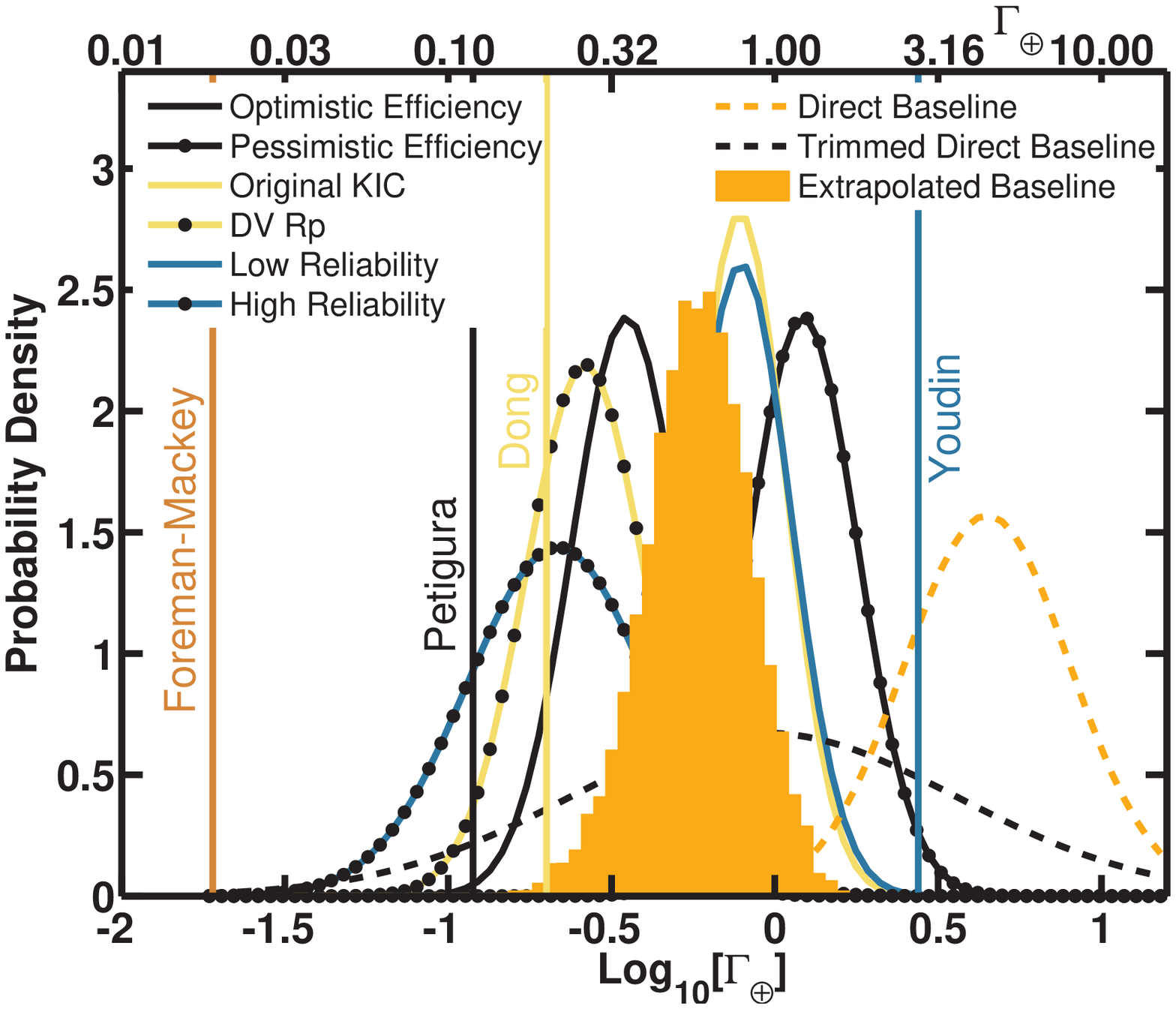}

\includegraphics[trim=0.25in 0.0in 0.35in 0.05in,scale=0.5,clip=true]{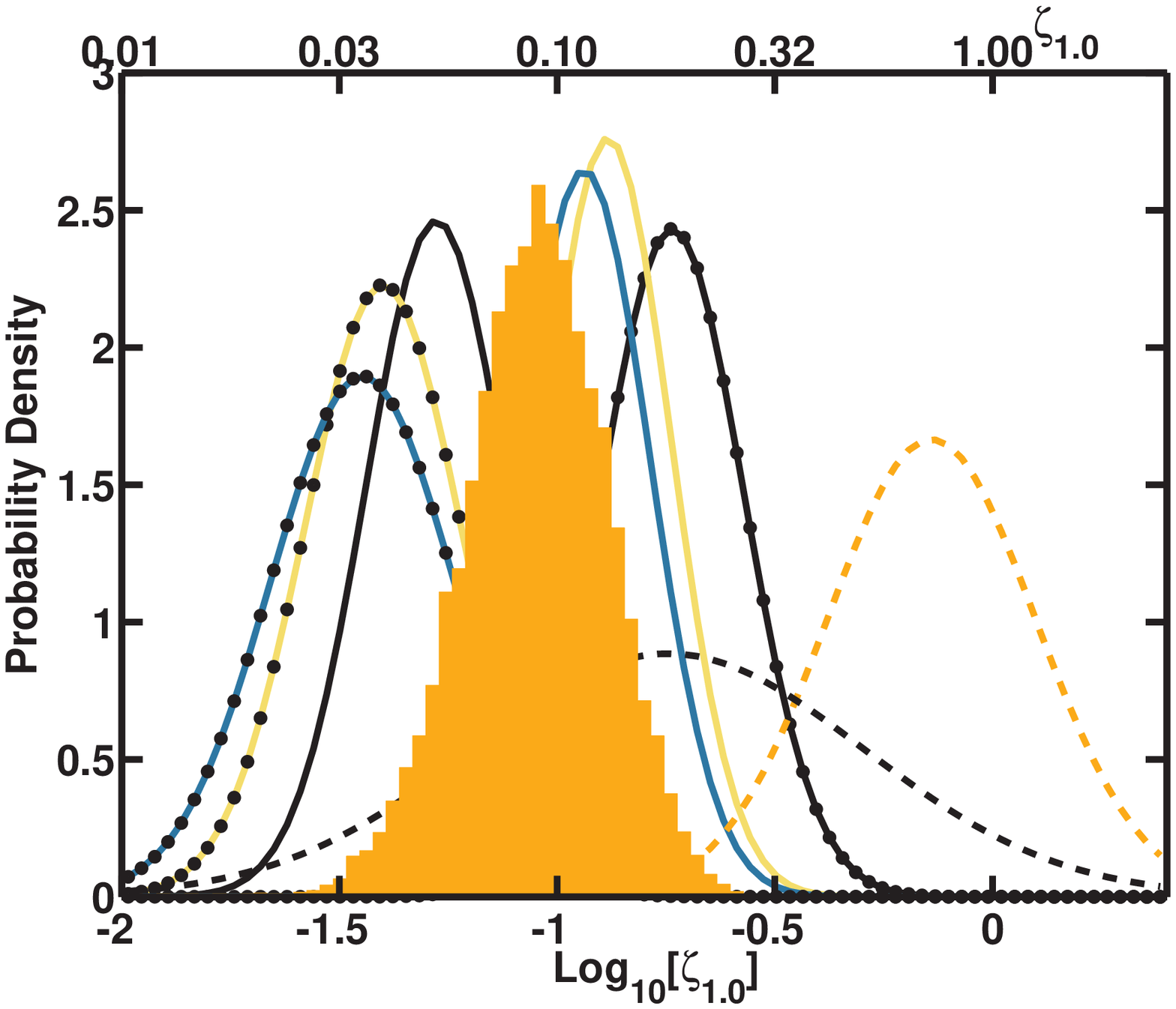}
\caption{
Top: distribution for the PLDF evaluated at the \Rp\ and \Porb\ of Earth, $\Gamma_{\oplus}$, using the extrapolated baseline analysis (filled orange histogram).  Solid curves represent results using alternative inputs with the same line types as in Figure~\ref{fig:syscomp}.  We also show two alternative analyses that directly measure, without extrapolation over \Porb, $\Gamma_{\oplus}$ from the Q1-Q16 \kepler\ planet candidate sample.  The direct measurement of $\Gamma_{\oplus}$ using the full long period planet candidate sample (orange dash curve) and the trimmed long period planet candidate sample (black dash curve) result in higher $\Gamma_{\oplus}$ than the extrapolated PLDF results, respectively. Previous $\Gamma_{\oplus}$ determinations from \citet{FOR14}, \citet{PET13B}, \citet{DON13}, and \citet{YOU11} are shown with vertical lines as labeled.  Bottom: same as top, but for the one year terrestrial planet occurrence rate, $\zeta_{1.0}$.\label{fig:earth}}
\end{figure}

First, we measure the PLDF evaluated at 1.0 \Rear\ and
\Porb=365.25~days,
$\Gamma_{\oplus}=dN/d\ln$\Porb$d\ln$\Rp\ \citep{YOU11,FOR14}.  In the
top panel of Figure~\ref{fig:earth}, we show the baseline (filled
orange histogram) $\Gamma_{\oplus}$ determined by extrapolating the
PLDF models from the 50$<$\Porb$<$300~days results.  We also show the
alternative systematic effects discussed in Section~\ref{sec:sensit}
(solid curves).  Note that the logarithmic scaling for the abscissa
indicates substantial systematic uncertainty in the results due to the
extrapolation.  We also show results for an extrapolated one year
terrestrial planet occurrence rate, $\zeta_{1.0}$, defined as the
occurrence rate of a planet within 20\% of the Earth's radius and
\Porb\ in the bottom panel of Figure~\ref{fig:earth}, for the baseline
(filled orange histogram) and alternative systematic effects discussed
in Section~\ref{sec:sensit} (solid curves).  For clarity the effects
of eccentricity and for `first planet only' are not displayed in
Figure~\ref{fig:earth} as the results are within the statistical
uncertainty of the extrapolated baseline result.

\subsection{Directly Measured At One Year Orbital Period}

Though the level of systematics present in our analysis are substantial
towards longer periods, we repeat the PLDF parameter
estimation in the 0.75$<$\Rp$<$2.5 \Rear\ and 300$<$\Porb$<$700~days
parameter space.  We show the average pipeline completeness for the
long period parameter space in Figure~\ref{fig:longpavgdet}.  The
planet candidates from the Q1-Q16 catalog of \citet{MULL15} are shown
as orange points and are indicated by a binary flag in
Table~\ref{tab:koiused}.  The analysis yields a high 
$F_{0}=4.7\pm^{3.1}_{1.77}$ planets per star, significantly steeper
$\alpha_{\rm avg}=-4.02\pm0.8$ and shallower $\beta=0.92\pm0.8$, where
the errors are the statistical uncertainty alone.

We defer a more detailed analysis of the systematics to future
work, but as a first step we consider culling the planet candidate
sample of the five planet candidates along the 0.01 pipeline
completeness contour shown in Figure~\ref{fig:longpavgdet}.  As
discussed in Section~\ref{sec:extrap}, this cluster of five planet
candidates represents a significant overabundance of planet candidates
relative to our shorter period analysis.  The overabundance relative
to the shorter period extrapolation prediction is nearly erased
(1.5-$\sigma$ overabundance), if the cluster of five planet candidates
is removed from the sample.  The KOIs belonging to the trimmed long
period planet candidate sample are indicated by a binary flag in
Table~\ref{tab:koiused}.  The PLDF parameter estimation after removing
these five planet candidates yields $F_{0}=1.7\pm^{1.2}_{0.6}$,
$\alpha_{\rm avg}=-2.7\pm1.1$, and $\beta=0.4\pm0.8$.  From this
direct analysis we show the one year terrestrial planet occurrence
rate in the bottom panel of Figure~\ref{fig:earth} for
$\zeta_{1.0}=0.76\pm^{0.55}_{0.33}$ (orange dash line) and
$\zeta_{1.0}=0.21\pm^{0.28}_{0.15}$ (black dash line), evaluated using
the full and clipped \kepler\ planet candidate sample in the
300$<$\Porb$<$700~days parameter space, respectively.

\subsection{One Year Terrestrial Planet Occurrence Rate Summary}

The wide range of occurrence rates obtained from this study is a
consequence of the difficulties associated with extrapolating, small
number statistics, and systematics (including false alarm
reliabilities).  This will impact refining $\zeta_{1.0}$ and HZ
statistics in future studies of the \kepler\ data set.  Compiling our
results of the extrapolated and direct analyses, we find
$\zeta_{1.0}=0.1$ with an allowed range of
0.01$\leq\zeta_{1.0}\leq$2.  Dynamical simulations cannot rule out
an upper limit of $\zeta_{1.0}\leq$2 \citep{SMI09}.  The mutual
hill radii separation for a system of three \Mp=1~\Mear\ planets
within the $\zeta_{1.0}$ occurrence region of a G~dwarf is $\gtrsim$9
corresponding to $\sim10^{10}$~yr stability \citep{SMI09}.  However,
for a lower mass K dwarf host and larger (\Rp=1.2\Rear) planets the
mutual hill radii separation $\sim 7$ for a triple planet system in
the $\zeta_{1.0}$ zone would likely be unstable on a $10^9$ yr timescale.

For the PLDF value at the \Rp\ and \Porb\ of Earth, we find
$\Gamma_{\oplus}=0.6$ with an acceptable range from
0.04$\leq\Gamma_{\oplus}\leq$11.5.  For comparison with previous
studies, we show in the bottom panel of Figure~\ref{fig:earth} as
vertical lines estimates of $\Gamma_{\oplus}$ from \citet{FOR14},
\citet{PET13B}, \citet{DON13}, and \citet{YOU11} from left to right,
respectively.  {\cjb In order to calculate results for
  $\Gamma_{\oplus}$ from the \citet{DON13} study, we extrapolate their
  parametric power law model as given for the
  1$\leq$\Rp$\leq$2~\Rear\ analysis from their Table~2.}  The
central value for $\Gamma_{\oplus}$ from \citet{FOR14} is in tension
with our analysis, but there is overlap in the upper tail of their
posterior with our lower limits.  {\cjb The analysis of \citet{FOR14}
  used the same inputs from \citet{PET13B}.  However, \citet{FOR14}
  determine that finding a steeper fall off of occurrence rates toward
  longer \Porb\ than \citet{PET13B} and taking into account
  uncertainty on planet radii lead to a systematically lower value for
  $\Gamma_{\oplus}$ than \citet{PET13B} when starting from the same
  inputs.  Further work is needed in order to isolate whether the
  differences between \citet{FOR14} and our study results
  predominately from differing inputs or methodology.}  The other
results for $\Gamma_{\oplus}$ from the literature are consistent with
our allowed range of $\Gamma_{\oplus}$.

\section{Conclusion}\label{sec:conclusion}

In this study we make use of the first \kepler\ pipeline run using
nearly all (Q1-Q16) the extant \kepler\ data in order to measure the
planet occurrence rate for 0.75$\leq$\Rp$\leq$2.5~\Rear in the
50$\leq$\Porb$\leq$300~days range orbiting the GK dwarf
\kepler\ sample.  We employ the first characterization of the
\kepler\ pipeline detection efficiency calibrated with transit
injection and recovery tests \citep{CHR15}, the Q1-Q16 \kepler\ planet
candidate catalog \citep{MULL15}, and the KIC stellar parameter
catalog revision of \citet{HUB14}.

We fit the observed planet candidate sample using a parametric PLDF
model following the work of \citet{YOU11} and explore alternative
inputs into the calculation in order to study the systematic errors
present.  In general, we find higher occurrence rates for the
mini-Neptune to terrestrial planet regime orbiting GK dwarfs and also
larger uncertainties driven by the systematics than indicated by
previous studies \citep{DON13,SIL15,PET13A,PET13B,MUL15}.  We
determine that $F_{0}=0.77$ planets per GK dwarf in the
\kepler\ sample have a planet within the
0.75$\leq$\Rp$\leq$2.5~\Rear\ and 50$\leq$\Porb$\leq$300~days regime
with a systematic dominated allowable range of 0.28$\leq F_{0}\leq$1.9.
The power law exponent for the \Rp\ dependence in the
PLDF model has a best value $\alpha_{\rm avg}$=-1.5 indicating an
increasing planet occurrence towards small planets, but the allowed
range, -3.25$\leq\alpha_{\rm avg}\leq$0.53, implies that we cannot
definitively determine whether the occurrence increases or decreases
towards the smallest planets.  However, fitting a double power-law
model over a wider range of 0.75$\leq$\Rp$\leq$5.0~\Rear\ does find
conclusive evidence for a break in the occurrence rate at $R_{\rm
  brk}=3.3\pm^{+0.2}_{-0.4}$.

We estimate a one year terrestrial planet occurrence rate,
$\zeta_{1.0}=0.1$, with an acceptable range
0.01$\leq\zeta_{1.0}\leq$2, by integrating within 20\% of the \Rp\ and
\Porb\ of Earth.  The narrower $\zeta_{1.0}$ parameter space is a
subset of the G~dwarf HZ, $\eta_{\oplus}$
\citep{KAS93,SEL07,ZSO13,KOP13}.  Thus, $\zeta_{1.0}$ places a lower
limit on $\eta_{\oplus}$ for G~dwarfs.  $\zeta_{1.0}$, which depends
upon the direct observable \Porb, facilitates comparison with future
\kepler\ occurrence rate studies and is readily compared to
theoretical terrestrial planet formation models.  We defer estimates
of $\eta_{\oplus}$, which depends upon the indirect observable of
stellar insolation and uncertain atmospheric evolution theory for
terrestrial planets outside the Solar System, to future studies.

We also determine a 0.6 year terrestrial planet occurrence rate,
$\zeta_{0.6}$=0.075, with an acceptable range
0.013$\leq\zeta_{0.6}\leq$0.30, by integrating within 20\% of the
\Rp=1~\Rear\ and \Porb$_{\venus}$ corresponding to Venus-type planets
for G~dwarf hosts.  For the K~dwarfs of our sample (\Teff$<$5200~K),
\Porb=0.6~yr roughly corresponds to the Solar-Earth insolation flux
level.  Thus, $\zeta_{0.6}$ places a lower limit on $\eta_{\oplus}$ for
K~dwarfs.

Although the current results are dominated by systematic
uncertainties, which, unlike statistical uncertainties that are
limited by the quantity and quality of data, can be minimized with
additional study.  From our analysis, we identify the leading sources
of systematics: instrumental false alarm contamination of the
planet candidate sample, determining planet radii (independent of the
degeneracy with \Rstar), pipeline completeness, and stellar
parameters.  Additional work on third light contamination, orbital
eccentricity, astrophysical false positives, and false negative rate
of the planet vetting process is needed.  All of these should be
examined carefully before accepting a definitive value for
$\eta_{\oplus}$.

\acknowledgments

We thank the referee for insightful suggestions which improved the manuscript.
Funding for this Discovery mission is provided by NASA's Science
Mission Directorate.  This research has made use of the NASA Exoplanet
Archive, which is operated by the California Institute of Technology,
under contract with the National Aeronautics and Space Administration
under the Exoplanet Exploration Program.  D.H. acknowledges support by
the Australian Research Council's Discovery Projects funding scheme
(project number DE140101364) and support by the National Aeronautics
and Space Administration under Grant NNX14AB92G issued through the
Kepler Participating Scientist Program.

\end{document}